\documentstyle{livrev}

\RequirePackage{epsf}
\RequirePackage{longtable}

\begin{document}

\title{Gravitational Waves from Gravitational Collapse}

\author{Kimberly C. B. New \\
        Los Alamos National Laboratory \\
        MS B220, X-2 \\
        Los Alamos, New Mexico 87545 \\
        e-mail:knew@lanl.gov \\
\\
\small{(last modified: 09 January 2003)}
}

\date{}
\maketitle

\begin{abstract}
Gravitational wave emission from the gravitational collapse of massive stars has been studied for more than three decades.  Current state of the art numerical investigations of collapse include those that use progenitors with more realistic angular momentum profiles, properly treat microphysics issues, account for general relativity, and examine non--axisymmetric effects in three dimensions.  Such simulations predict that gravitational waves from various phenomena associated with gravitational collapse could be detectable with ground--based and space--based interferometric observatories.
\end{abstract}


\newpage


\section{Introduction}
\label{section:introduction}

The field of gravitational wave (GW) astronomy will soon become a reality. The first generation of ground--based interferometric detectors (LIGO~\cite{ligo}, VIRGO~\cite{virgo}, GEO 600~\cite{geo600}, TAMA 300~\cite{tama300}) are beginning their search for GWS.  Towards the end of this decade, two of these detectors (LIGO, VIRGO) will undergo upgrades that should allow them to reach sensitivities necessary to regularly detect emission from astrophysical sources.  A space--based interferometric detector, LISA~\cite{lisa}, could be launched in the early part of the next decade. One important class of sources for these observatories is stellar gravitational collapse.  This class includes the accretion induced collapse (AIC) of white dwarf binary components and the core collapse of massive stars ($M>8 M_{\odot} $), very massive Population III stars ($M$=$100$-$500 M_{\odot} $), and supermassive stars (SMSs, $M>10^6 M_{\odot}$).  Some of these collapses result in explosions (Type II, Ib/c supernovae and hypernovae) and all leave behind neutron star or black hole remnants.

Strong GWs can be emitted during a gravitational collapse/explosion and, following the collapse, by the resulting compact remnant~\cite{thorne-94, mueller-97a, mueller-98, finn-99, schutz-99, fryer-02, fryer-02b, hughes-02}.  GW emission during the collapse itself may result if the collapse or explosion involves aspherical bulk mass motion or convection.  Rotational or fragmentation instabilities encountered by the collapsing star will also produces GWs.  Neutron star remnants of collapse may emit GWs due to the growth of rotational or r-mode instabilities.  Black hole remnants will also be sources of GWs if they experience accretion induced ringing.  All of these phenomena have the potential of being detected by gravitational wave observatories because they involve the rapid change of dense matter distributions.

Observation of gravitational collapse by gravitational wave detectors will provide unique information, complementary to that derived from electromagnetic and neutrino detectors.
Gravitational radiation arises from the coherent superposition of mass motion.  Whereas, electromagnetic emission is produced by the incoherent superposition of radiation from electrons, atoms, and molecules.  Thus GWs carry different kinds of information than other types of radiation.  Furthermore, electromagnetic radiation interacts strongly with matter and thus only gives a view of the collapse from lower density regions near the surface of the star and is weakened by absorption as it travels to the detector.  Neutrinos can escape from much further within the collapsing star, but even they are scattered by the highest density regions in the core. By contrast, gravitational waves can propagate from the innermost parts of the stellar core to detectors without attenuation by intervening matter.

The characteristics of the GW emission from gravitational collapse have been the subject of much study.  Core collapse supernovae, in particular, have been investigated as sources of gravitational radiation for more than three decades (see, e.g.,~\cite{ruffini-71, thuan-74, saenz-78, detweiler-81, nakamura-81, mueller-82, stark-85, finn-90, monchmeyer-91, zwerger-97, rampp-98, fryer-02, fryer-02b}). However, during this time research has produced estimates of GW strength that vary over orders of magnitude.  This is due to the complex nature of core collapse. Important theoretical and numerical issues include 
\begin{itemize}
\item construction of accurate progenitor models, including realistic angular momentum distributions,
\item proper treatment of microphysics, including the use of realistic equations of state and neutrino transport,
\item simulation in three-dimensions to study non-axisymmetric effects,
\item inclusion of general relativistic effects, 
\item inclusion of magnetic field effects, and
\item study of the effect of an envelope on core behavior.
\end{itemize}
To date, collapse simulations generally include state of the art treatments of only one or two of the above physics issues (often because of numerical constraints).  For example, those studies
that include advanced microphysics have often been run with Newtonian gravity (and approximate evaluation of the GW emission; see section~\ref{section:aic-num}).  A 3D, general relativistic collapse simulation, which includes all significant physics effects, is not feasible at present.  However, good progress has been made on the majority of the issues listed above; the more recent work will be reviewed in some detail here.

The remainder of this article is structured as follows. Each category of gravitational collapse will be discussed in a separate section (AIC in section~\ref{section:aic}, collapse of massive stars in section~\ref{section:sne}, collapse of Population III stars in section~\ref{section:pop3}, and collapse of SMSs in section~\ref{section:sms}).  Each of these sections (\ref{section:aic}, \ref{section:sne}, \ref{section:pop3}, \ref{section:sms}) is divided into subsection topics: Collapse Scenario, Formation Rate, GW Emission Mechanisms, and Numerical Predictions of GW Emission.  In the subsections on numerical predictions, the detectability of the GW emission from various phenomena associated with collapse is examined.  In particular, the predicted characteristics of GW emission are compared to the sensitivities of LIGO (for sources with frequencies of $1$ to $10^4\,{\rm Hz}$) and LISA (for sources with lower frequencies in the range of $10^{-4}$ to $1\,{\rm Hz}$).
    
\newpage


\section{Accretion Induced Collapse}
\label{section:aic}

\subsection{Collapse Scenario}
\label{subsection:aic-scen}

As a white dwarf accretes matter from a binary companion, its density will increase.  If the rate of the resulting compressional heating exceeds the white dwarf's cooling rate, the star's temperature will increase as well. When the mass of the white dwarf exceeds the Chandrasekhar stability limit, it will begin to collapse.

If the temperature of the core is high enough, nuclear burning will begin and the stellar
pressure will increase.  However, electron capture will act to reduce both the temperature
and pressure behind the burn front.  The collapse outcome is determined by the relative
strength of the nuclear burning and electron capture.  If nuclear burning is strong enough,
the white dwarf will explode as a Type Ia supernova (SN).  If electron capture wins out, the
collapse will continue and result in neutron star formation.  Electron capture will dominate if the density at which nuclear ignition occurs exceeds a critical density $\rho_{crit}$.
For C-O white dwarfs, $\rho_{crit}$ is in the range $ 6 \times 10^9-10^{10}\, {\rm g\, cm}^{-3} $
\cite{bravo-99}. For O-Ne-Mg white dwarfs, electron capture may be stronger than nuclear
burning under most conditions (if a Rayleigh-Taylor instability does not produce a turbulent burn front)~\cite{nomoto-91, isern-91}.  The collapse of an O-Ne-Mg white dwarf begins when its central density reaches $4 \times 10^9\,{\rm g cm}^{-3}$.  (For more details about the conditions under which AIC occurs, see~\cite{nomoto-91, isern-91, bravo-99, brachwitz-00, liu-01}.) 

The dynamics of the collapse itself are somewhat similar to the dynamics of core collapse SNe (without the resulting explosive disruption of the star).  The collapse is halted when nuclear densities are reached in the core.  The core bounces and sends a bounce--shock outward through
the star.  After several milliseconds, the shock stalls and the remainder of the star collapses through the shock front (see, e.g.,~\cite{woosley-92}). Less than $10^{-1} M_{\odot}$ will likely be ejected by the star (due to the bounce itself or to neutrino absorption/wind mechanisms)~\cite{fryer-99}.

\subsection{Formation Rate}
\label{section:aic-form}

The AIC occurrence rate is difficult to determine for a number of reasons.  These include incomplete understanding of binary star evolution and accretion processes~\cite{Cassisi-98, Kato-99}.  Another uncertainty is whether the collapse of an accreting Chandrasekhar mass white dwarf results in a supernovae explosion or a complete AIC (with accompanying neutron star formation).

The AIC rate can be indirectly inferred from the observed amount of rare, neutron rich isotopes present in the Galaxy.  These isotopes (formed via electron capture) are present in the portion ($\sim 0.1 M_{\odot}$) of the outer envelope ejected by the star during an AIC.  If all of these isotopes present in the Galaxy are assumed to have originated in AICs, an upper limit of $\sim 10^{-5} {\rm yr}^{-1}$ can be set for the galactic AIC rate~\cite{fryer-99}.

Binary population synthesis analysis can be used to determine which accreting white dwarfs will undergo AIC.  The results of Yungelson and Livio~\cite{yungelson-98} predict that the galactic AIC rate is between $8 \times 10^{-7}$ and $8 \times 10^{-5} {\rm yr}^{-1}$.  Thus, a reasonable occurrence rate can be found for an observation distance of $100\,{\rm Mpc}$.

\subsection{GW Emission Mechanisms}
\label{section:aic-mech}

During AIC, emission of GWs will occur if the infall of matter is aspherical.  GWs will also be produced if the collapsing star or neutron star remnant encounters non--axisymmetric instabilities.
These include global rotational mode, $r$--mode, and fragmentation instabilities.

Global rotational instabilities in fluids arise from non--axisymmetric modes $e^{\pm im\phi}$, where $m$=$2$ is known as the ``bar--mode'' \cite{tassoul-78, andersson-02}. It is convenient to parameterize a system's susceptibility to these modes by the stability parameter $\beta=T_{rot}/|W|$.  Here $T_{rot}$ is the rotational kinetic energy and $W$ is the gravitational potential energy. Dynamical rotational instabilities, driven by Newtonian hydrodynamics and gravity, develop on the order of the rotation period of the object.  For the uniform density, incompressible, uniformly rotating Maclaurin spheroids, the dynamical bar-mode instability sets in at $\beta_d \approx 0.27$. For differentially rotating fluids with a polytropic equation of state, numerical simulations have determined that the stability limit $\beta_d \approx 0.27$ is valid for initial angular momentum distributions that are similar to those of Maclaurin spheroids~\cite{shapiro-83, durisen-85, managan-85, imamura-85, pickett-96, houser-98, toman-98}.  If the object has an off-center density maximum, $\beta_d$ could be as low as $0.10$~\cite{tohline-90, woodward-94, pickett-96, centrella-01}. Furthermore, recent work by Shibata, Kariono, and Eriguchi~\cite{shibata-02c} suggests that $\beta_d$ could be as low as $\sim 0.01$ for stars with a large degree of differential rotation.  General relativity may enhance the dynamical bar--mode instability by slightly reducing $\beta_d$~\cite{shibata-00, saijo-01}. Secular rotational instabilities are driven by dissipative processes such as gravitational radiation reaction and viscosity.  When this type of instability arises, it develops on the timescale of the relevant dissipative mechanism, which can be much longer than the rotation period.  The secular bar--mode instability limit for Maclaurin spheroids is $\beta_s \approx 0.14$.

In rotating stars, gravitational radiation reaction drives the $r$--modes towards unstable growth~\cite{andersson-98, friedman-98}. In hot, rapidly rotating neutron stars, this instability may not be suppressed by internal dissipative mechanisms (such as viscosity and magnetic fields)~\cite{lindblom-98}.  If not limited, the dimensionless amplitude $\alpha$ of the dominant ($m$=$2$) $r$--mode will grow to order unity within ten minutes of the formation of a neutron star rotating with a millisecond period.  The GWs emitted carry away angular momentum, and will cause the newly formed neutron star to spin down over time.  The spindown timescale and the strength of the GWs themselves are directly dependent on the maximum value $\alpha_{max}$ to which the amplitude is allowed to grow~\cite{lindblom-01, lindblom-02}.  Originally, it was thought that $\alpha_{max}\sim 1$.  Later work indicated that $\alpha_{max}$ may be $\geq 3$~\cite{lindblom-01, stergioulas-01, schenk-02, lindblom-02}.  Some research suggests that magnetic fields, hyperon cooling, and hyperon bulk viscosity may limit the growth of the $r$--mode instability, even in nascent neutron stars~\cite{jones-01a, jones-01b, rezzolla-01a, rezzolla-01b, lindblom-02, lindblom-02a, haensel-02, andersson-02} (significant uncertainties remain regarding the efficacy of these dissipative mechanisms).  Furthermore, a recent study of a simple barotropic neutron star model by Arras et al.~\cite{arras-02} suggests that multimode couplings could limit $\alpha_{max}$ to values $\ll 1$.   If $\alpha_{max}$ is indeed $\ll 1$, GW emission from $r$--modes in collapsed remnants is likely undetectable.  For the sake of completeness, an analysis of GW emission from $r$--modes (which assumes $\alpha_{max} \sim 1$)  is presented in the remainder of this paper.  However, because it is quite doubtful that $\alpha_{max}$ is sizeable, $r$--mode sources are omitted from figures comparing source strengths and detector sensitivities and from discussions of likely detectable sources in the concluding section.

There is some numerical evidence that a collapsing star may fragment into two or more orbiting clumps~\cite{fryer-01}.  If this does indeed occur, the orbiting fragments would be a strong GW source.

\subsection{Numerical Predictions of GW Emission}
\label{section:aic-num}

\subsubsection{Full Collapse Simulations}

The accretion--induced collapse of white dwarfs has been simulated by a number of groups~\cite{baron-87, mayle-88, woosley-92, fryer-99}.  The majority of these simulations have been Newtonian and have focused on mass ejection and neutrino and $\gamma$-ray emission during the collapse and its aftermath (note that neglecting relativistic effects likely introduces an error of order $(v^2/c^2) \sim 10\%$ for the neutron star remnants of AIC; see below).  The most sophisticated are those carried out by Fryer et al.~\cite{fryer-99}, as they include realistic equations of state, neutrino transport, and rotating progenitors. 

As a part of their general evaluation of upper limits to GW emission from gravitational collapse, Fryer, Holz, and Hughes (hereafter, FHH)~\cite{fryer-02} examined an AIC simulation (model 3) of Fryer et al.~\cite{fryer-99}. FHH used both numerical and analytical techniques to estimate the peak amplitude $h_{pk}$, energy $E_{GW}$, and frequency $f_{GW}$ of the gravitational radiation emitted during the collapse simulations they studied.

For direct numerical computation of the GWs emitted in these simulations, FHH used the quadrupole approximation, valid for nearly Newtonian sources~\cite{misner-73}.  This approximation is standardly used to compute the GW emission in Newtonian simulations.  The reduced or traceless quadrupole moment of the source can be expressed as
\def\ibar{{\hbox{\rlap{\hbox{\raise.25ex\hbox{-}}}I\llap{\hbox{\raise.25ex\hbox{-}}}}}}
\def\dddibar{\stackrel{\scriptscriptstyle \cdot\cdot\cdot}{{\ibar}}_{ij}}
\begin{equation}
\ibar_{ij} = \int \rho(x_i x_j - \frac{1}{3} \delta_{ij} r^2)d^3r,
\label{equation:ibar}
\end{equation}
where $i,j$=$1,2,3$ are spatial indices and $r$=$(x^2+y^2+z^2)^{1/2}$ is the distance to the source.  The two polarizations of the gravitational wave field, $h_+$ and $h_{\times}$, can be computed in terms of $\dddibar$, where an overdot indicates a time derivative $d/dt$.
The energy $E_{GW}$ is a function of $\dddibar$.  Equation~(\ref{equation:ibar}) can be used to calculate $\ibar_{ij}$ from the results of a numerical simulation by direct summation over the computational grid. Numerical time derivatives of $\ibar_{ij}$ can then be taken to compute $h_+$, $h_{\times}$, and $E_{GW}$. However, successive application of numerical derivatives generally introduces artificial noise.  Methods for computing $\ddot{\ibar}_{ij}$ and $\dddibar$ without taking numerical time derivatives have been developed~\cite{finn-90, blanchet-90, monchmeyer-91}.  These methods recast the time derivatives of $\ibar_{ij}$ as spatial derivatives of hydrodynamic quantities computed in the collapse simulation (including the density, velocities, and gravitational potential) .  Thus, 
instantaneous values for $\ddot{\ibar}_{ij}$ and $\dddibar$ can be computed on a single numerical time slice (see~\cite{monchmeyer-91, zwerger-97} for details).  Note that FHH define $h_{pk}$ as the maximum value of the rms strain $h=\sqrt{\langle{h_{+}^2 + h_{\times}^2\rangle}}$, where angular brackets indicate that averages have been taken over both wavelength and viewing angle on the sky. 

Errors resulting from the neglect of general relativistic effects (in collapse evolutions as a whole and in GW emission estimations like the quadrupole approximation) are of order $v^2/c^2$.  These errors are typically $\sim 10\%$ for neutron star remnants of AIC, $<30\%$ for neutron star remnants of massive stellar collapse, and $>30\%$ for black hole remnants.  Neglect of general relativity in rotational collapse studies is of special concern because relativistic effects counteract the stabilizing effects of rotation (see section~\ref{section:sne-num}).

Because the code used in the collapse simulations examined by FHH~\cite{fryer-99} was axisymmetric, their use of the numerical quadrupole approximation discussed above does not account for GW emission that may occur due to non--axisymmetric mass flow.  The GWs computed directly from their simulations come only from polar oscillations (which are significant when the mass flow during collapse [or explosion] is largely aspherical).  

In order to predict the GW emission produced by non--axisymmetric instabilities, FHH employed rough analytical estimates.  The expressions they used to approximate the rms strain $h$ and the power $P=dE_{GW}/dt$ of the GWs emitted by a star that has encountered the bar--mode instability are
\begin{equation}
h_{bar} = \sqrt{\frac{32}{45}} \frac{G}{c^4}\frac{m r^2 \omega^2}{d}
\label{equation:hbar}
\end{equation}
and
\begin{equation}
P_{bar} = \frac{32}{45} \frac{G}{c^5} m^2 r^4 \omega^6.
\label{equation:pbar}
\end{equation}
Here $m$, $2r$, $\omega$ are the mass, length, and angular frequency of the bar and $d$ is the distance to the source.  FHH vary the mass $m$ assumed to be enclosed by the bar (which has a corresponding length $2r$) and compute the characteristics of the GW emission as a function of this enclosed mass.
For simplicity, FHH assumed that a fragmentation instability will cause a star to break into two clumps (although more clumps could certainly be produced). Their estimates for the rms strain and power radiated by the orbiting binary fragments are
\begin{equation}
h_{bin} = \sqrt{\frac{128}{5}} \frac{G}{c^4}\frac{m r^2 \omega^2}{d}
\label{equation:hbin}
\end{equation}
and
\begin{equation}
P_{bin} = \frac{128}{5} \frac{G}{c^5} m^2 r^4 \omega^6.
\label{equation:pbin}
\end{equation}
Here $m$ is the mass of a single fragment, $2r$ is the separation of the fragments, $\omega$ is their orbital frequency.

For their computation of the GWs radiated via r-modes, FHH used the method of Ho and Lai~\cite{ho-00} (which assumes $\alpha_{max}$=$1$) and calculated only the emission from the dominant $m$=$2$ mode. This approach is detailed in FHH. If the neutron star mass and initial radius are taken to be $1.4 M_{\odot}$ and $12.53\,{\rm km}$, respectively, the resulting formula for the average GW strain is
\begin{equation}
h(t) = 1.8 \times 10^{-24} \alpha \left(\frac{\nu_s}{1 {\rm kHz}} \right) \left(\frac{20 {\rm Mpc}}{d} \right),
\label{equation:hrmode}
\end{equation}
where $\alpha$ is the mode amplitude and $\nu_s$ is the spin frequency.

FHH's numerical quadrupole estimate of the GWs from polar oscillations in the AIC simulation of Fryer et al.~\cite{fryer-99}, predicts a peak dimensionless amplitude $h_{pk}=5.9\times 10^{-24}$ (for $d$=$100\,{\rm Mpc}$).  The energy $E_{GW} = 3 \times 10^{45} {\rm ergs}$ is emitted at a frequency of $f_{GW} \approx 50\,{\rm Hz}$.  This amplitude is about an order of magnitude too small to be observed by the advanced LIGO-II detector. The sensitivity curve for the broadband configuration of the LIGO-II detector is shown in figure ~\ref{figure:ligo} (see appendix A of~\cite{fryer-02} for details on the computation of this curve). Note that the characteristic strain $h$ is plotted along vertical axis in figure~\ref{figure:ligo} (and in LISA's sensitivity curve, shown in figure~\ref{figure:lisa}).  For burst sources, $h=h_{pk}$.  For sources which persist for $N$ cycles, $h=\sqrt{N}h_{pk}$.

\begin{figure}[h]
 \def\epsfsize#1#2{0.65#1}
 \centerline{\epsfbox{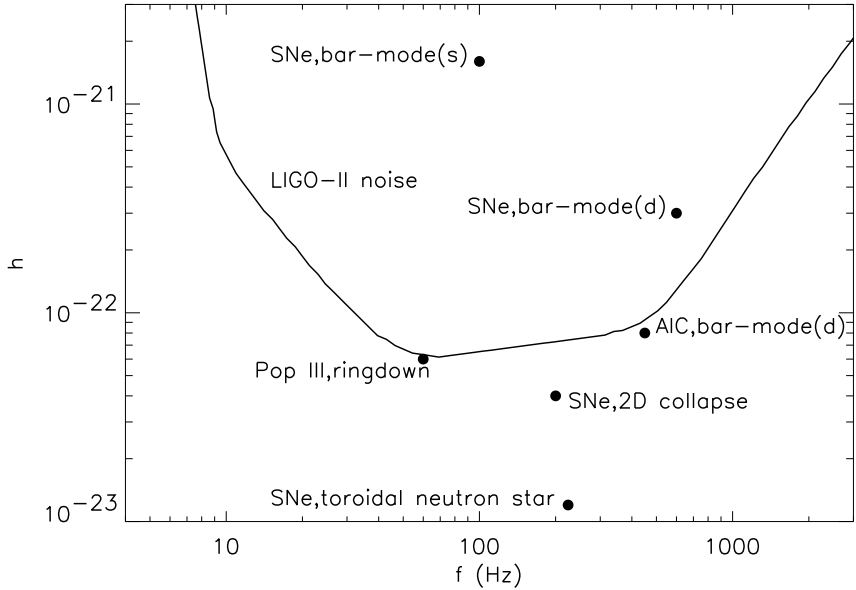}}
 \caption{\it A comparison between the GW amplitude $h(f)$ for various sources and the LIGO-II sensitivity curve. See the text for details regarding the computations of $h$.  The AIC sources are assumed to be located at a distance of $100\,{\rm Mpc}$; the SNe sources at $10\,{\rm Mpc}$; and the Population III sources at a luminosity distance of $\sim 50\,{\rm Gpc}$.  Secular bar--mode sources are identified with an (s), dynamical bar--modes with a (d).}
 \label{figure:ligo}
\end{figure}

The simulation of Fryer et al.~\cite{fryer-99} considered by FHH does produce an object with an off-center density maximum. However, because the maximum $\beta$ reached in the AIC simulation of Fryer et al.\ was $<0.06$ (and because the degree of differential rotation present in the remnant was low~\cite{fryer-02p}), the collapsing object is not likely to encounter dynamical rotational instabilities.
The choice of initial angular momentum $J$ for the progenitor in an AIC simulation can affect this outcome.  Fryer et al.'s choice of $J$=$10^{49} {\rm g cm^2 s^{-1}}$ is such that the period of the white dwarf progenitor is $10{\rm s}$ less than the shortest observed period $30 {\rm s}$ of a cataclysmic variable white dwarf~\cite{king-91}. If higher values of $J$ exist in accreting white dwarfs, bar--mode instabilities may be more likely to occur (see discussion of work by Liu and Lindblom below).

According to FHH, the remnant of this AIC simulation will be susceptible to $r$--mode growth.  Assuming $\alpha_{max} \sim 1$ (which is likely not physical; see section~\ref{section:aic-mech}), they predict $E_{GW} > 10^{52} {\rm ergs}$.  FHH compute $h(f_{GW})$ for coherent observation of the neutron star as it spins down over the course of a year. For a neutron star located at a distance of $100 {\rm Mpc}$, this track is always below the LIGO-II noise curve. The point on this track with the maximum $h$, which corresponds to the beginning of $r$--mode evolution, is shown in figure~\ref{figure:ligo}.  The track moves down and to the left (i.e., $h$ and $f$ decrease) in this figure as the $r$-mode evolution continues.

\subsubsection{Analyses of Equilibrium Representations of Collapse}
In addition to full hydrodynamics collapse simulations, many studies of gravitational collapse have used hydrostatic equilibrium models to represent stars at various stages in the collapse process.  Some investigators use sequences of equilibrium models to represent snapshots of the phases of collapse (e.g.,~\cite{baumgarte-99, new-01a, liu-01}).  Others use individual equilibrium models as initial conditions for hydrodynamical simulations(e.g.,~\cite{smith-96, pickett-96, new-00, centrella-01}).  Such simulations represent the approximate evolution of a model beginning at some intermediate phase during collapse or the evolution of a collapsed remnant.  These studies do not typically follow the intricate details of the collapse itself.  Instead, their goals include determining the stability of models against the development of non-axisymmetric modes and estimation of the characteristics of any resulting GW emission. 

Liu and Lindblom~\cite{liu-01, liu-02} have applied this equilibrium approach to AIC.  Their investigation began with a study of equilibrium models built to represent neutron stars formed from AIC~\cite{liu-01}.  These neutron star models were created via a two step process, using a Newtonian version of Hachisu's self-consistent field method~\cite{hachisu-86}.  Hachisu's method ensures that the forces due to the centrifugal and gravitational potentials and the pressure are in balance in the equilibrium configuration. 

Their process of building the nascent neutron stars began with the construction of rapidly rotating, pre--collapse white dwarf models. Their Models I and II are C-O white dwarfs with central densities $\rho_c$=$10^{10}$ and $6 \times 10^9  {\rm g\,cm}^{-3}$, respectively (recall this is the range of densities for which AIC is likely for C-O white dwarfs).  Their Model III is an O-Ne-Mg white dwarf that has $\rho_c=4 \times 10^9  {\rm g\,cm}^{-3}$ (recall this is the density at which collapse is induced by electron capture).  All three models are uniformly rotating, with the maximum allowed angular velocities.  The models' values of total angular momentum are roughly $3$--$4$ times that of Fryer et al.'s~\cite{fryer-99} AIC progenitor Model 3.  The realistic equation of state used to construct the white dwarfs is a Coulomb corrected, zero temperature, degenerate gas equation of state~\cite{salpeter-61, chandra-67}.

In the second step of their process, Liu and Lindblom~\cite{liu-01} built equilibrium models of the collapsed neutron stars themselves.  The mass, total angular momentum, and specific angular momentum distribution of each neutron star remnant is identical to that of its white dwarf progenitor (see section 3 of~\cite{liu-01} for justification of the specific angular momentum conservation assumption).  These models were built with two different realistic neutron star equations of state.   

Their cold neutron star remnants had values of the stability parameter $\beta$ ranging from $0.23$-$0.26$.  It is interesting to compare these results with those of Zwerger and M\"{u}ller~\cite{zwerger-97}.  Zwerger and M\"{u}ller performed axisymmetric hydrodynamics simulations of stars with polytropic equations of state ($P\propto \rho^{\Gamma}$).  Their initial models were $\Gamma$=$4/3$ polytropes, representative of massive white dwarfs. All of their models started with $\rho_c$=$10^{10} {\rm g\,cm}^{-3}$.  Their model that was closest to being in uniform rotation (A1B3) had 22\% less total angular momentum than Liu and Lindblom's Model I.  The collapse simulations of Zwerger and M\"{u}ller that started with model A1B3 all resulted in remnants with values of $\beta < 0.07$.  Comparison of the results of these two studies could indicate that the equation of state may play a significant role in determining the structure of collapsed remnants.  Or it could suggest that the assumptions employed in the simplified investigation of Liu and Lindblom are not fully appropriate.  Zwerger and M\"{u}ller's work will be discussed in much more detail in section~\ref{section:sne}, as it was performed in the context of core collapse supernovae.  

In a continuation of the work of Liu and Lindblom, Liu~\cite{liu-02} used linearized hydrodynamics to perform a stability analysis of the cold neutron star AIC remnants of Liu and Lindblom~\cite{liu-01}. He found that only the remnant of the O-Ne-Mg white dwarf (Liu and Lindblom's Model III) developed the dynamical bar--mode ($m$=$2$) instability.  This model had an initial $\beta$=$0.26$.
Note that the $m$=$1$ mode, observed by others to be the dominate mode in unstable models with values of $\beta$ much lower than $0.27$~\cite{tohline-90, woodward-94, pickett-96, centrella-01}, did not grow in his simulation.  Because Liu and Lindblom's Models I and II had
lower values of $\beta$, Liu identified the onset of instability for neutron stars formed via AIC as $\beta_d\approx 0.25$.

Liu estimated the peak amplitude of the GWs emitted by the Model III remnant to be $h_{pk}\approx 1.4 \times 10^{-24}$ and the LIGO-II signal to noise ratio (for a persistent signal like that seen in the work of~\cite{new-00} and~\cite{brown-00}) to be $S/N \leq 3$ (for $f_{GW}\approx 450\,{\rm Hz}$. These values are for a source located at $100 {\rm Mpc}$.  He also predicted that the timescale for gravitational radiation to carry away enough angular momentum to eliminate the bar--mode is $\tau_{GW} \sim 7 {\rm s}$ ($\sim 3\times 10^3$ cycles).  Thus, $h\sim 8 \times 10^{-23}$. (Note this value for $h$ is merely an upper limit as it assumes that the amplitude and frequency of the GWs do not change over the $7\,{\rm s}$ during which they are emitted. They will of course change as angular momentum is carried away from the object via GW emission.) Such a signal may be marginally detectable with LIGO--II (see figure~\ref{figure:ligo}).  Details of the approximations on which these estimates are based can be found in~\cite{liu-02}. 

Liu cautions that his results hold if the magnetic field of the proto--neutron star is $B\leq 10^{12}\,{\rm G}$.  If the magnetic field is larger, then it may have time to suppress some of the neutron star's differential rotation before it cools.  This would make bar formation less likely.  Such a large field could only result if the white dwarf progenitor's $B$ field was $\geq 10^8\,{\rm G}$.  Observation based estimates suggest that about $25\%$ of white dwarfs in interacting close binaries (cataclysmic variables) are magnetic and that the field strengths for these stars are $\sim 10^{7}$--$3\times 10^8\,{\rm G}$~\cite{wickramasinghe-00}.

\subsection{Going Further}
\label{section:aic-gf}

The AIC scenario is generally discussed in terms of the collapse of an accreting white dwarf to a neutron star.  However, Shibata, Baumgarte, and Shapiro have examined the collapse of a rotating, supramassive neutron star to a black hole~\cite{shibata-00a}.  Such supramassive neutron stars (with masses greater than the maximum mass for a nonrotating neutron star) could be formed and pushed to collapse via accretion from a binary companion.  They performed 3D, fully general relativistic hydrodynamics simulations of uniformly rotating neutron stars.  Dynamical non--axisymmetric instabilities (such as the bar--mode) did not have time to grow in their simulations prior to black hole formation.  Differentially rotating neutron star progenitors could have higher values of $\beta$ than the uniformly rotating models used in this study and may be susceptible to non--axisymmetric instabilities on a shorter timescale.

\newpage


\section{Collapse of Massive Stars}
\label{section:sne}

\subsection{Collapse Scenario}
\label{subsection:sne-scen}

\def\gtrsim{\stackrel{\scriptstyle >}{\scriptstyle \sim}}
\def\lesssim{\stackrel{\scriptstyle <}{\scriptstyle \sim}}
Stars with mass greater than $\sim 8 M_{\odot}$ will undergo core collapse at the end of their thermonuclear burning life cycles.  The collapse will trigger a supernova explosion if the star's mass is $<40-50 M_{\odot}$ (for reviews, see~\cite{arnett-89, bethe-90, burrows-00}).  Core collapse SNe include Types II and Ib/Ic.  SNe Ib/Ic are distinguished from Type II SNe by the absence of hydrogen in their spectra.  SNe Ib/Ic are thought to result from the collapse of the cores of massive stars that have lost their hydrogen envelopes (and possibly part of their helium envelopes) by stellar winds or by mass transfer.  The SNe Ib/Ic progenitors that lose their outer envelopes via stellar winds are known as Wolf-Rayet stars and have initial masses $\gtrsim 30 M_{\odot}$; those that undergo mass loss via mass transfer in binaries have progenitor masses in the range $12 M_{\odot}$--$18 M_{\odot}$~\cite{mueller-98}.

It is believed that the remnant of a core collapse SN is a neutron star if the mass of the progenitor is less than $\sim 20-25 M_{\odot}$~\cite{fryerkalo-01, fryer-02}.  If the progenitor's mass is in the range $20-25 \lesssim M \lesssim 40-50 M_{\odot}$, the entire star is not ejected in the SNe explosion. More than $2 M_{\odot}$ will fall back onto the nascent neutron star and lead to black hole formation.  If the progenitor's mass exceeds roughly $40-50 M_{\odot}$, then no explosion will occur and the star will collapse directly to a black hole.  These objects are known as collapsars.  However, it is unclear if high metallicity stars with $M \gtrsim 40-50 M_{\odot}$ actually end their lives in collapse or are prevented from doing so by stellar wind driven mass loss.~\cite{fryer-02} Note that the limits on the progenitor masses quoted in this paragraph (especially the $40-50 M_{\odot}$ lower limit for direct black hole formation) are uncertain because the progenitor mass dependence of the neutrino explosion mechanism (see below) is unknown~\cite{hamuy-02, nadyozhin-02}.

The massive iron cores of SNe II/Ib/Ic progenitors are supported by both thermal and electron degeneracy pressures.  The density and temperature of such a core will eventually rise, due to the build up of matter consumed by thermonuclear burning, to the point where electron capture and photodissociation of nuclei begin.  Dissociation lowers the photon and electron temperatures and thereby reduces the core's thermal support~\cite{finn-89}.  Electron capture reduces the electron degeneracy pressure. One or both of these processes will trigger the collapse of the core.  The relative importance of dissociation and electron capture in instigating collapse is determined by the mass of the star~\cite{finn-89}.  The more massive the core, the bigger the role played by dissociation.

Approximately $70\%$ of the inner portion of the core collapses homologously and subsonically.  The outer core collapses at supersonic speeds~\cite{finn-89, monchmeyer-91}.  The maximum velocity of the outer regions of the core reaches $\sim 7 \times 10^4\, {\rm km\,s^{-1}}$.  It takes just $1\,{\rm s}$ for an earth--sized core to collapse to a radius of $50\,{\rm km}$~\cite{arnett-89}.

The inward collapse of the core is halted by nuclear forces when its central density $\rho_c$ is $2$--$10$ times the density of nuclear material~\cite{baron-85a, baron-85b}.   The core overshoots its equilibrium position and bounces.  A shock wave is formed when the supersonically infalling outer layers hit the rebounding inner core.  If the inner core pushes the shock outwards with energy $E>10^{51} {\rm ergs}$ (supplied by the binding energy of the nascent neutron star), then the remainder of the star can be ejected in about $20\,{\rm ms}$~\cite{arnett-89}.  This so-called ``prompt explosion'' mechanism will only succeed if a very soft supra-nuclear equation of state is used in conjunction with a relatively small core ($M \lesssim 1.35 M_{\odot}$, derived from a very low mass progenitor) and a large portion of the collapse proceeds homologously~\cite{bethe-90, mueller-98}.  Inclusion of general relativistic effects in collapse simulations can increase the success of the prompt mechanism in some cases~\cite{baron-85a, swesty-94}.

Both dissociation of nuclei and electron capture can reduce the ejection energy, causing the prompt mechanism to fail.  The shock will then stall at a radius in the range $100$--$200 {\rm km}$.  Colgate and White~\cite{colgate-66} were the first to suggest that energy from neutrinos emitted by the collapsed core could revive the stalled shock.  (See Burrows and Thompson~\cite{burrows-02} for a recent review of core collapse neutrino processes.) However, their simulations did not include the physics necessary to accurately model this ``delayed explosion'' mechanism.  Wilson and collaborators were the first to perform collapse simulations with successful delayed ejections~\cite{bowers-82a, bowers-82b, wilson-85, bethe-85, wilson-86, wilson-88}.  However, their simulations and those of others had difficulty producing energies high enough to match observations~\cite{colgate-89, bruenn-93, janka-93}.

Observations of SN 1987a show that significant mixing occurred during this supernovae (see Arnett et al.~\cite{arnett-89} for a review).  Such mixing can be attributed to nonradial motion resulting from fluid instabilities. Convective instabilities play a significant role in the current picture of the delayed explosion mechanism. The outer regions of the nascent neutron star are convectively unstable after the shock stalls (for an interval of $10$-$100 {\rm ms}$ after bounce) due to the presence of negative lepton and energy gradients~\cite{mueller-98}.  This has been confirmed by both 2D and 3D simulations~\cite{burrows-92, burrows-93, janka-93, mueller-93, mueller-94, janka-96a, mueller-98, mezzacappa-98, janka-01, janka-02c, rampp-02}.  Convective motion is more effective at transporting neutrinos out of the proto--neutron star than is diffusion.  Less than $10\%$ of the neutrinos emitted by the neutron star need to be absorbed and converted to kinetic energy for the shock to be revived~\cite{mueller-98}.  The ``hot bubble'' region above the surface of the neutron star has also been shown to be convectively unstable~\cite{colgate-89, bethe-90, colgate-93, fryer-02a}.  Janka and M\"{u}ller have demonstrated that convection in this region only aids the explosion if the neutrino luminosity is in a narrow region~\cite{janka-96b}.
Some simulations that include advanced neutrino transport methods have cast doubt on the ability of convection to ensure the success of the delayed explosion mechanism~\cite{mezzacappa-98b}.  However, Rampp and Janka~\cite{janka-02c, rampp-02} have recently performed collapse simulations with sophisticated neutrino physics (including Boltzmann transport and state of the art neutrino--matter interactions) that are very close to producing explosions.  Their results indicate that successful explosions could be very sensitive to the details of the neutrino physics. Multi-dimensional simulations that include the full effects of general relativity, rotation, and an improved treatment of neutrino interactions will likely be necessary to properly model the delayed explosion mechanism~\cite{janka-01a, janka-02c}.

In addition to the mixing seen in SN 1987a, observations of polarization in the spectra of several core collapse SNe, jets in the Cas A remnant, and kicks in neutron stars suggest that supernovae are inherently aspherical (see~\cite{arnett-89, akiyama-02, hoflich-02} and references therein).  (Note that these asphericities could originate in the central explosion mechanism itself and/or the mechanism[s] for energy transfer between the core and ejecta~\cite{janka-02c}.)  These observations partly motivated the multidimensional studies of convection in the delayed explosion mechanism discussed above and have also spurred investigations of the role jets may play in SNe explosions~\cite{macfayden-99, wheeler-00, woosley-02}.  Recent 3D radiation hydrodynamics simulations performed by H\"{o}flich et al.~\cite{hoflich-02} indicate that low velocity jets stalled inside SN envelopes can account for the observed asymmetries.  Possible sources of such asymmetries include magnetic fields~\cite{baumgarte-02, akiyama-02}, anisotropic neutrino emission, and convection.

\subsection{Formation Rate}
\label{section:sne-form}

Type II/Ib/Ic supernovae are observed to occur in only spiral and irregular galaxies.  The most thorough computation of SNe rates is that of Cappellaro et al.~\cite{cappellaro-99}.  Their sample includes 137 SNe from five different SN searches.  They determined that the core collapse SNe rate in the Galaxy is $0.6 \times 10^{-3}$--$1.6 \times 10^{-2} {\rm yr}^{-1}$. Thus, a reasonable occurrence rate can be found for an observation distance of $10\,{\rm Mpc}$. (Note that a recent infrared survey estimates that the rate may actually be an order of magnitude higher~\cite{maiolino-02}.) Approximately $5-40\%$ of these SNe will leave behind black hole remnants~\cite{fryerkalo-01}.  The formation rate of collapsars, the massive cores that collapse to black holes without an accompanying SNe explosion, is unknown.  This is because of the uncertainty in stellar wind--driven mass--loss rates~\cite{fryerkalo-01}. 

\subsection{GW Emission Mechanisms}
\label{section:sne-mech}

Gravitational radiation will be emitted during the collapse/explosion of a core collapse SNe due to the star's changing quadrupole moment.  A rough description of the possible evolution of the quadrupole moment is given in the remainder of this paragraph. During the first $100$--$250 \,{\rm ms}$ of the collapse, as the core contracts and flattens, the magnitude of the quadrupole moment $\ibar_{ij}$ will increase.  The contraction speeds up over the next $20\,{\rm ms}$ and the density distribution becomes a centrally condensed torus~\cite{monchmeyer-91}.  In this phase the core's shrinking size dominates its increasing deformation and the magnitude of $\ibar_{ij}$ decreases.  As the core bounces, $\ibar_{ij}$ changes rapidly due to the deceleration and rebound.  If the bounce occurs because of nuclear pressure, its timescale will be $< 1 {\rm ms}$.  If centrifugal forces play a role in halting the collapse, the bounce can last up to several ${\rm ms}$~\cite{monchmeyer-91}. The magnitude of $\ibar_{ij}$ will increase due to the core's expansion after bounce.  As the resulting shock moves outwards, the unshocked portion of the core will undergo oscillations, causing $\ibar_{ij}$ to oscillate as well.  The shape of the core, the depth of the bounce, the bounce timescale, and the rotational energy of the core all strongly affect the GW emission.  For further details see~\cite{finn-89, monchmeyer-91}.

Convectively driven inhomogeneities in the density distribution of the outer regions of the nascent neutron star and anisotropic neutrino emission are other sources of GW emission during the collapse/explosion~\cite{burrows-96, mueller-97}.

As discussed in the case of AIC, global rotational instabilities (such as the $m$=$2$ bar--mode) may develop during the collapse itself or in a neutron star remnant.  A neutron star remnant will likely also be susceptible to the radiation reaction driven $r$-modes.  Both of these types of instabilities will emit GWs, as will a fragmentation instability if one occurs.  See section~\ref{section:aic-mech} for further details regarding these instabilities. 

If the collapsed remnant is a black hole, GWs will be radiated as the infall of the remaining stellar matter distorts the black hole's geometry.  This ``ringdown phase'' will end when gravitational radiation has dissipated all of the black hole's accretion-induced distortion.
Zanotti, Rezzolla, and Font~\cite{zanotti-02} have recently suggested that the torus of matter surrounding the black hole may be an even stronger source of GWs than the collapse itself (see also~\cite{kobayashi-02, putten-02}).  When perturbed, such ``toroidal neutron stars'' may undergo regular oscillations and thus emit copious GWs.

\subsection{Numerical Predictions of GW Emission}
\label{section:sne-num}

\subsubsection{Historical Investigations}

The collapse of the progenitors of core collapse supernovae has been investigated as a source of gravitational radiation for more than three decades.  In an early study published in 1971, Ruffini and Wheeler~\cite{ruffini-71} identified mechanisms related to core collapse that could produce GWs and provided order of magnitude estimates of the characteristics of such emission.

Quantitative computations of GW emission during the infall phase of collapse were performed by Thuan and Ostriker~\cite{thuan-74} and Epstein and Wagoner~\cite{epstein-75, epstein-76}, who simulated the collapse of oblate dust spheroids.  Thuan and Ostriker used Newtonian gravity and computed the emitted radiation in the quadrupole approximation.  Epstein and Wagoner discovered that post-Newtonian effects prolonged the collapse and thus lowered the GW luminosity.
Subsequently, Novikov~\cite{novikov-75} and Shapiro and Saenz~\cite{shapiro-77, saenz-78} included internal pressure in their collapse simulations and were thus able to examine the GWs emitted as collapsing cores bounced at nuclear densities.  The quadrupole GWs from the ringdown of the collapse remnant were initially investigated by the perturbation study of Turner and Wagoner~\cite{turner-79} and later by Saenz and Shapiro~\cite{saenz-79,saenz-81}.

M\"{u}ller~\cite{mueller-82} calculated the quadrupole GW emission from 2D axisymmetric collapse based on the Newtonian simulations of M\"{u}ller and Hillebrandt~\cite{mueller-81} (these simulations used a realistic equation of state and included differential rotation).  He found that differential rotation enhanced the efficiency of the GW emission.

The first fully general relativistic investigation of stellar core collapse were 
Nakamura's 2D simulations of rotating collapse~\cite{nakamura-81,nakamura-83}. However, because of the limits of his numerical formalism and computational resources, he was unable to compute the emitted gravitational radiation (the energy of this emission is quite small compared to the rest mass energy and thus was difficult to extract numerically).  The results of this work indicate that collapse does not lead to black hole formation if the parameter $q=J/M^2$ exceeds unity (here $J$ and $M$ are the angular momentum and gravitational mass of the remnant).

Stark and Piran~\cite{stark-85,piran-86} were the first to compute the GW emission from fully relativistic collapse simulations, using the ground--breaking formalism of Bardeen and Piran~\cite{bardeen-83}.  They followed the (pressure--cut induced) collapse of rotating polytropes in 2D. Their work focused in part on the conditions for black hole formation and the nature of the resulting ringdown waveform (which they found could be described by the quasi--normal modes of a rotating black hole).  In each of their simulations, less than $1\%$ of the gravitational mass was converted to GW energy.

Seidel and collaborators also studied the effects of general relativity on the GW emission during collapse and bounce~\cite{seidel-87, seidel-88}.  They employed a perturbative approach, valid only in the slowly rotating regime.

The gravitational radiation from non--axisymmetric collapse was investigated by Detweiler and Lindblom who used a sequence of non--axisymmetric ellipsoids to represent the collapse evolution~\cite{detweiler-81}.  They found that the radiation from their analysis of non--axisymmetric collapse was emitted over a more narrow range of frequency than in previous studies of axisymmetric collapse.

For further discussion of the first two decades of study of the GW emission from stellar collapse see~\cite{finn-91}.  In the remainder of section~\ref{section:sne-num}, more recent investigations will be discussed.

\subsubsection{Axisymmetric Simulations}

The core collapse simulations of M\"{o}nchmeyer et al.\ began with better iron core models and a more realistic microphysics treatment (including a realistic equation of state, electron capture, and a simple neutrino transport scheme) than any previous study of GW emission from axisymmetric stellar core collapse~\cite{monchmeyer-91}.  (The shortcomings of their investigation included initial models that were not in rotational equilibrium, an equation of state that was somewhat stiff in the subnuclear regime, and use of Newtonian gravity.)  Each of their four models had a different initial angular momentum profile.  The rotational energies of the models ranged from $0.1$--$.45$ of the maximum possible rotational energy.

The collapses of three of the four models of M\"{o}nchmeyer et al.\ were halted by centrifugal forces at subnuclear densities.  This type of low $\rho_c$ bounce had been predicted by Shapiro and Lightman~\cite{shapiro-76} and Tohline~\cite{tohline-84} (in the context of the ``fizzler'' scenario for failed supernovae; see also~\cite{hayashi-98, hayashi-99, imamura-01}) and had been observed in earlier collapse simulations~\cite{mueller-80, symbalisty-84}.  M\"{o}nchmeyer and collaborators found that a bounce caused by centrifugal forces would last for several ${\rm ms}$, whereas a bounce at nuclear densities would occur in $<1\,{\rm ms}$.  They also determined that a subnuclear bounce produced larger amplitude oscillations in density and radius, with larger oscillation periods, than a bounce initiated by nuclear forces alone.  They pointed out that these differences in timescale and oscillatory behavior should affect the GW signal.  Therefore, the GW emission could indicate whether the bounce was a result of centrifugal or nuclear forces. 

\begin{figure}[h]
 \def\epsfsize#1#2{0.4#1}
 \centerline{\epsfbox{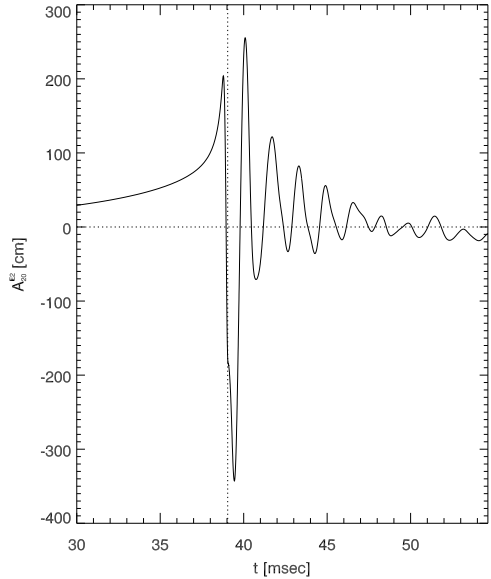}}
 \caption{\it Type I waveform (quadrupole amplitude $A^{E2}_{20}$ as a function of time) from one of Zwerger and M\"{u}ller's~\cite{zwerger-97} simulations of a collapsing polytrope. The vertical dotted line marks the time at which the first bounce occurred. (Figure 5d of ~\cite{zwerger-97}; used with permission.)}
 \label{figure:typeI}
\end{figure}

\begin{figure}[h]
 \def\epsfsize#1#2{0.4#1}
 \centerline{\epsfbox{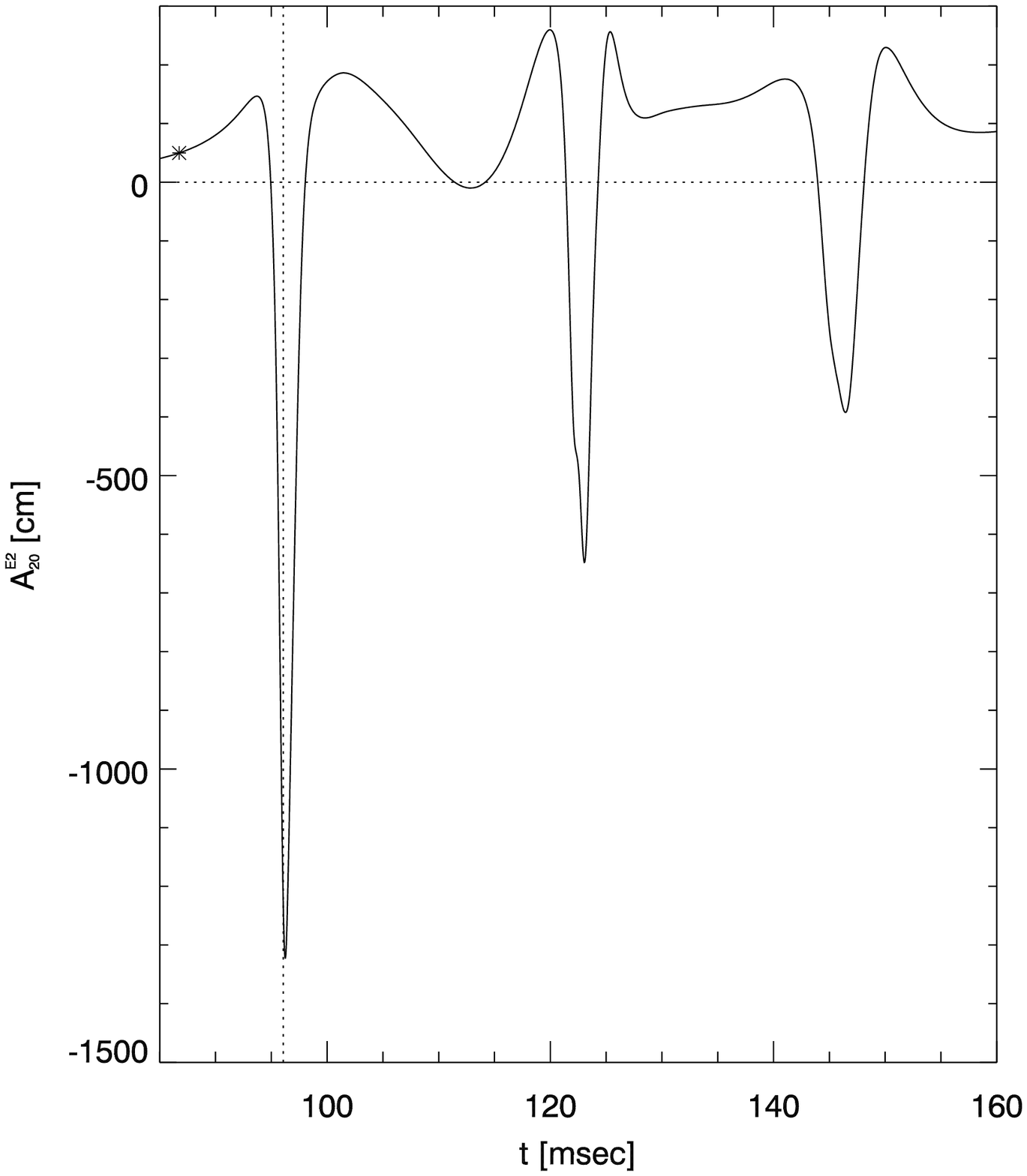}}
  \caption{\it Type II waveform (quadrupole amplitude $A^{E2}_{20}$ as a function of time) from one of Zwerger and M\"{u}ller's~\cite{zwerger-97} simulations of a collapsing polytrope. The vertical dotted line marks the time at which bounce occurred. (Figure 5a of ~\cite{zwerger-97}; used with permission.)}
 \label{figure:typeII}
\end{figure}

M\"{o}nchmeyer et al.\ identified two different types of waveforms in their models (computed using the numerical quadrupole approximation discussed in section~\ref{section:aic-num}).  The waveforms they categorized as Type I (similar to those observed in previous collapse simulations~\cite{mueller-82,finn-90}) are distinguished by a large amplitude peak at bounce and subsequent damped ringdown oscillations.  They noted that Type I signals were produced by cores that bounced at nuclear densities (or bounces at subnuclear densities if the cores had small ratios of radial kinetic to rotational kinetic energies).  The quadrupole gravitational wave amplitude $A^{E2}_{20}$ for a Type I waveform is shown in figure~\ref{figure:typeI} (see~\cite{thorne-80, zwerger-97} for expressions relating $A^{E2}_{20}$ to $h$). The waveforms identified as Type II exhibit several maxima, which result from multiple bounces.  See figure~\ref{figure:typeII} for an example of a Type II waveform.  Note that the waveforms displayed in figures~\ref{figure:typeI},~\ref{figure:typeII} are from the study of Zwerger and M\"{u}ller~\cite{zwerger-97}, discussed below.

The model of M\"{o}nchmeyer et al.\ that bounced due to nuclear forces had the highest GW amplitude of all of their models, $h_{pk} \sim 10^{-23}$ for a source distance $d=10\,{\rm Mpc}$, and largest emitted energy $E_{GW} \sim 10^{47}\,{\rm erg}$.  The accompanying power spectrum peaked in the frequency range $5\times 10^2$--$10^3\,{\rm Hz}$.

The most extensive Newtonian survey of the parameter space of axisymmetric, rotational core collapse is that of Zwerger and M\"{u}ller~\cite{zwerger-97}.  They simulated the collapse of 78 initial models with varying amounts of rotational kinetic energy (reflected in the initial value of the stability parameter $\beta_i$), differential rotation, and equation of state stiffness.  In order to make this large survey tractable, they used a simplified equation of state and did not explicitly account for electron capture or neutrino transport.  Their initial models were constructed in rotational equilibrium via the method of Eriguchi and M\"{u}ller~\cite{eriguchi-85}.  The models had a polytropic equation of state, with initial adiabatic index $\Gamma_i=4/3$.  Collapse was induced by reducing the adiabatic index to a value $\Gamma_r$ in the range $1.28$--$1.325$.  The equation of state used during the collapse evolution had both polytropic and thermal contributions.

\begin{figure}[h]
 \def\epsfsize#1#2{0.4#1}
 \centerline{\epsfbox{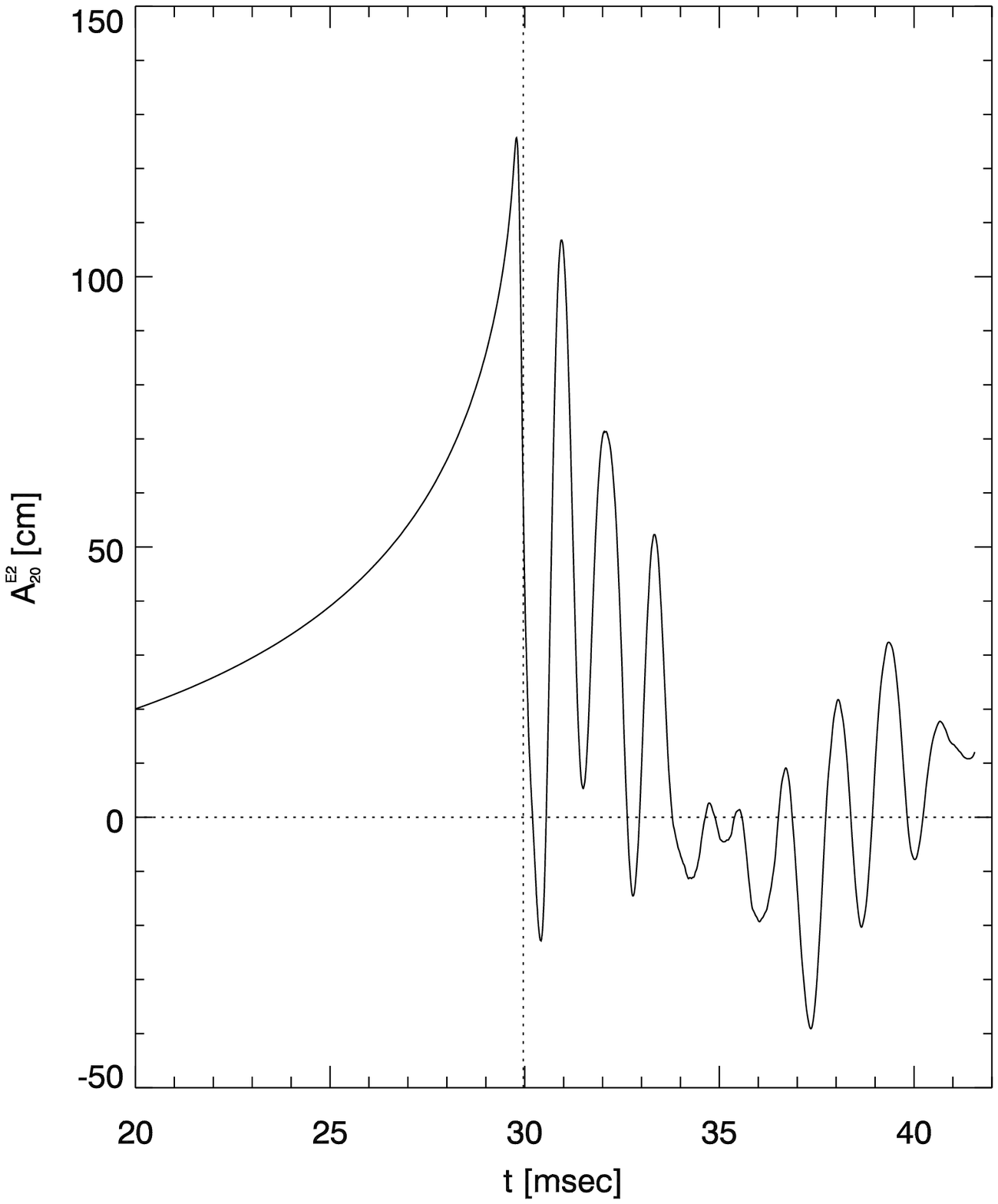}}
  \caption{\it Type III waveform (quadrupole amplitude $A^{E2}_{20}$ as a function of time) from one of Zwerger and M\"{u}ller's~\cite{zwerger-97} simulations of a collapsing polytrope. The vertical dotted line marks the time at which bounce occurred. (Figure 5e of ~\cite{zwerger-97}; used with permission).}
 \label{figure:typeIII}
\end{figure}

The major result of Zwerger and M\"{u}ller's investigation was that the signal type of the emitted gravitational waveform in their runs was determined by the stiffness of the equation of state of the collapsing core (i.e., the value of $\Gamma_r$).  In their simulations, Type I signals (as labelled by M\"{o}nchmeyer et al.~\cite{monchmeyer-91}) were produced by models with relatively soft equations of state $\Gamma_r \lesssim 1.31$.  Type II signals were produced by the models with stiffer equations of state ($\Gamma_r \gtrsim 1.32$).  They found a smooth transition between these signal types if $\Gamma_r$ was increased while all other parameters were held fixed.  They also observed another class of signal, Type III, for their models with the lowest $\Gamma_r$=$1.28$.  Type III waveforms have a large positive peak just prior to bounce, a smaller negative peak just after bounce, and smaller subsequent oscillations with very short periods (see figure~\ref{figure:typeIII}). Type III signals were not observed in the evolution of strongly differentially rotating $\Gamma_r$=$1.28$ models.  Their waveforms were computed with the same technique used in~\cite{monchmeyer-91}.

In contrast to the results of M\"{o}nchmeyer et al.~\cite{monchmeyer-91}, in Zwerger and M\"{u}ller's investigation, the value of $\rho_c$ at bounce did not determine the signal type.  Instead, the only effect on the waveform due to $\rho_c$ was a decrease in $h_{pk}$ in models that bounced at subnuclear densities.  The effect of the initial value of $\beta_i$ on $h_{pk}$ was non-monotonic.  For models with $\beta_i \lesssim 0.1$, $h_{pk}$ increased with increasing $\beta_i$.  This is because the deformation of the core is larger for faster rotators.  However, for models with larger $\beta_i$, $h_{pk}$ decreases as $\beta_i$ increases.  These models bounce at subnuclear densities.  Thus the resulting acceleration at bounce, and GW amplitude, are smaller.  Zwerger and M\"{u}ller found that the maximum value of $h_{pk}$ for a given sequence was reached when $\rho_c$ at bounce was just less than $\rho_{nuc}$.  The degree of differential rotation did not have a large effect on the emitted waveforms computed by Zwerger and M\"{u}ller.  However, they did find that models with soft equations of state emitted stronger signals as the degree of differential rotation increased.  

The models of Zwerger and M\"{u}ller that produced the largest GW signals fell into two categories: those with stiff equations of state and $\beta_i < .01$; and those with soft equations of state, $\beta_i \geq .018$, and large degrees of differential rotation.  The GW amplitudes emitted during their simulations were in the range $4 \times 10^{-25} \lesssim h \lesssim 4 \times 10^{-23}$, for $d$=$10\,{\rm Mpc}$ (the model with the highest $h$ is identified in figure~\ref{figure:ligo}).  The corresponding energies ranged from $10^{44} \lesssim E_{gw} \lesssim 10^{47}\,{\rm erg}$.  The peaks of their power spectra were between $500\,{\rm Hz}$ and $1\,{\rm kHz}$.  Such signals would fall just outside of the range of LIGO--II.

Yamada and Sato~\cite{yamada-95} used techniques very similar to those of Zwerger and M\"{u}ller~\cite{zwerger-97} in their core collapse study.  Their investigation revealed that the $h_{pk}$ for Type I signals saturated when the dimensionless angular momentum of the collapsing core, $q=J/(2GM/c)$, reached $\sim 0.5$. They also found that $h_{pk}$ was sensitive to the stiffness of the equation of state for densities just below $\rho_{nuc}$.  The characteristics of the GW emission from their models were similar to those of Zwerger and M\"{u}ller.

\subsubsection{Non--Axisymmetric Simulations}

The GW emission from non--axisymmetric hydrodynamics simulations of stellar collapse was first  studied by Bonazzola and Marck~\cite{marck-92, bonazzola-93}. They used a Newtonian, pseudo-spectral hydrodynamics code to follow the collapse of polytropic models.  Their simulations covered only the pre--bounce phase of the collapse.  They found that the magnitudes of $h_{pk}$ in their 3D simulations were within a factor of two of those from equivalent 2D simulations and that the gravitational radiation efficiency did not depend on the equation of state. 

The first 3D hydrodynamics collapse simulations to study the GW emission well beyond the core bounce phase were performed by Rampp, M\"{u}ller, and Ruffert~\cite{rampp-98}.  These authors started their Newtonian simulations with the only model (A4B5G5) of Zwerger and M\"{u}ller~\cite{zwerger-97} that had a post-bounce value of the stability parameter $\beta$=$0.35$ that significantly exceeded $0.27$ (recall this is the value at which the dynamical bar instability sets in for Maclaurin spheroid--like models).  This model had the softest EOS ($\Gamma_r$=$1.28$), highest $\beta_i$=$0.04$, and largest degree of differential rotation of all of Zwerger and M\"{u}ller's models.  The model's initial density distribution had an off-center density maximum (and therefore a torus-like structure).  Rampp, M\"{u}ller, and Ruffert evolved this model with a 2D hydrodynamics code until its $\beta$ reached $\approx 0.1$.  At that point, $2.5\,{\rm ms}$ prior to bounce, the configuration was mapped onto a 3D nested cubical grid structure and evolved with a 3D hydrodynamics code.  

Before the 3D simulations started, non--axisymmetric density perturbations were imposed to seed the growth of any non--axisymmetric modes to which the configuration was unstable.  When the perturbation imposed was random ($5\%$ in magnitude), the dominant mode that arose was $m$=$4$.  The growth of this particular mode was instigated by the cubical nature of the computational grid.  When an $m$=$3$ perturbation was imposed ($10\%$ in magnitude), three clumps developed during the post-bounce evolution and produced three spiral arms. These arms carried mass and angular momentum away from the center of the core.  The arms eventually merged into a bar-like structure (evidence of the presence of the $m$=$2$ mode).  Significant non--axisymmetric structure was visible only within the inner $40\,{\rm km}$ of the core.  Their simulations were carried out to $\sim 14\,{\rm ms}$ after bounce.

The amplitudes of the emitted gravitational radiation (computed in the quadrupole approximation) were only $\sim 2 \%$ different from those observed in the 2D simulation of Zwerger and M\"{u}ller.  Because of low angular resolution in the 3D runs, the energy emitted was only $65\%$ of that emitted in the corresponding 2D simulation.  

The findings of Centrella et al.~\cite{centrella-01} indicate it is possible that some of the post-bounce configurations of Zwerger and M\"{u}ller, which have lower values of $\beta$ than the model studied by Rampp, M\"{u}ller, and Ruffert~\cite{rampp-98}, may also be susceptible to non-axisymmetric instabilities. Centrella et al.\ have performed 3D hydrodynamics simulations of $\Gamma$=$1.3$ polytropes to test the stability of configurations with off-center density maxima (as are present in many of the models of Zwerger and M\"{u}ller~\cite{zwerger-97}).  The simulations carried out by Centrella and collaborators were not full collapse simulations, but rather began with differentially rotating, equilibrium models.  These simulations tracked the growth of any unstable non--axisymmetric modes that arose from initial $1\%$ random density perturbations imposed.  Their results indicate that such models can become dynamically unstable at values of $\beta \gtrsim 0.14$.  The instability observed had a dominant $m$=$1$ mode.  Centrella et al.\ estimate that if a stellar core of mass $M\sim 1.4 M_{\odot}$ and radius $R\sim 200\,{\rm km}$ encountered this instability, the values of $h_{pk}$ from their models would be $\sim 2 \times 10^{-24}$--$2 \times 10^{-23}$, for $d$=$10\,{\rm Mpc}$.  The frequency at which $h_{pk}$ occurred in their simulations was $\sim 200\,{\rm Hz}$.  This instability would have to persist for at least $\sim 15$ cycles to be detected with LIGO--II.

Brown~\cite{brown-01} carried out an investigation of the growth of non--axisymmetric modes in post--bounce cores that was similar in many respects to that of Rampp, M\"{u}ller, and Ruffert~\cite{rampp-98}.  He performed 3D hydrodynamics simulations of the post--bounce configurations resulting from 2D simulations of core collapse.  His pre--collapse initial models are $\Gamma=4/3$ polytropes in rotational equilibrium.  The differential rotation laws used to construct Brown's initial models were motivated by the stellar evolution study of Heger, Langer, and Woosley~\cite{heger-00}.  The angular velocity profiles of their pre-collapse progenitors were broad and Gaussian-like.  Brown's initial models had peak angular velocities ranging from $0.8$--$2.4$ times those of~\cite{heger-00}.  The model evolved by Rampp, M\"{u}ller, and Ruffert~\cite{rampp-98} had much stronger differential rotation than any of Brown's models.  To induce collapse, Brown reduced the adiabatic index of his models to $\Gamma$=$1.28$, the same value used by~\cite{rampp-98}.

Brown found that $\beta$ increased by a factor $\lesssim 2$ during his 2D collapse simulations.  This is much less than the factor of $\sim 9$ observed in the model studied by Rampp, M\"{u}ller, and Ruffert~\cite{rampp-98}.  This is likely a result of the larger degree of differential rotation in the model of Rampp et al. 

Brown performed 3D simulations of the two most rapidly rotating of his post-bounce models (models $\Omega 24$ and $\Omega 20$, both of which had $\beta > 0.27$ after bounce) and of the model of Rampp et al.\ (which, although it starts out with $\beta$=$0.35$, has a sustained $\beta<0.2$). Brown refers to the Rampp et al.\ model as model RMR. Because Brown's models do not have off-center density maxima, they are not expected to be unstable to the $m$=$1$ mode observed by Centrella et al.~\cite{centrella-01}.  He imposed random $1\%$ density perturbations at the start of all three of these 3D simulations (note this perturbation was of a much smaller amplitude than those imposed by~\cite{rampp-98}).  

Brown's simulations determined that both his most rapidly rotating model $\Omega24$ (with post-bounce $\beta>0.35$) and model RMR are unstable to growth of the $m$=$2$ bar--mode. However, his model $\Omega20$ (with post-bounce $\beta>0.3$) was stable.  Brown observed no dominant $m$=$3$ or $4$ modes growing in model RMR at the times at which they were seen in the simulations of Rampp et al.  This suggests that the mode growth in their simulations was a result of the large perturbations they imposed.  The $m$=$2$ mode begins to grow in model RMR at about the same time as Rampp et al.\ stopped their evolutions.  No substantial $m$=$1$ growth was observed.

The results of Brown's study indicate that the overall $\beta$ of the post--bounce core may not be a good diagnostic for the onset of instability.  He found, as did Rampp, M\"{u}ller, and Ruffert~\cite{rampp-98}, that only the innermost portion of the core (with $\rho>10^{10}\,{\rm g\,cm^{-3}}$) is susceptible to the bar--mode.  This is evident in the stability of his model $\Omega20$. This model had an overall $\beta>0.3$ but an inner core with $\beta_{ic}=0.15$.  Brown also observed that the $\beta$ of the inner core does not have to exceed $0.27$ for the model to encounter the bar--mode.  Models $\Omega24$ and RMR had $\beta_{ic} \approx 0.19$.  He speculates that the inner cores of these later two models may be bar unstable because interaction with their outer envelopes feeds the instability or because $\beta_d < 0.27$ for such configurations.

Fryer and Warren~\cite{fryer-02a} have recently performed the first 3D collapse simulations to follow the entire collapse through explosion.  They used a smoothed particle hydrodynamics code,
a realistic equation of state, the flux--limited diffusion approximation for neutrino transport, and Newtonian spherical gravity. Their initial model was nonrotating. Thus, no bar--mode instabilities could develop during their simulations.  The only GW emitting mechanism present in their models was convection in the core.  The maximum amplitude $h$ of this emission, computed in the quadrupole approximation, was $\sim 3 \times 10^{-26}$, for $d$=$10\,{\rm Mpc}$~\cite{fryer-02b}.  Preliminary results from rotating simulations performed with this same 3D code do not show the development of bar or fragmentation instabilities~\cite{fryer-02b}.  

The GW emission from nonradial quasinormal mode oscillations in proto--neutron stars has recently been examined by Ferrari, Miniutti, and Pons~\cite{ferrari-02}.  They found that the frequencies of emission $f_{GW}$ during the first second after formation ($600$-$1100\,{\rm Hz}$ for the first fundamental and gravity modes) are significantly lower than the corresponding frequencies for cold neutron stars and thus reside in the bandwidths of terrestrial interferometers.  However, for first generation interferometers to detect the GW emission from an oscillating proto--neutron star located at $10\,{\rm Mpc}$, with a signal to noise ratio of $5$,
$E_{GW}$ must be$\sim 10^{-3}$--$10^{-2} M_{\odot}c^2$.  It is unlikely that this much energy is stored in these modes (the collapse itself may only emit $\sim 10^{-7} M_{\odot}c^2$ in gravitational waves~\cite{dimmelmeier-02b}).

\subsubsection{General Relativistic Simulations}

General relativistic effects oppose the stabilizing influence of rotation in pre--collapse cores.  Thus stars that might be prevented from collapsing due to rotational support in the Newtonian limit, may collapse when general relativistic effects are considered.  Furthermore, general relativity will cause rotating stars undergoing collapse to bounce at higher densities than in the Newtonian case~\cite{tassoul-78, zwerger-97, rampp-98, bruenn-01}.  

The full collapse simulations of Fryer and Heger~\cite{fryer-00} are the most sophisticated axisymmetric simulations from which the resultant GW emission has been studied~\cite{fryer-02, fryer-02b}.  Fryer and Heger include the effects of general relativity, but assume (for the purposes of their gravity treatment only) that the mass distribution is spherical.  The GW emission from these simulations was evaluated with either the quadrupole approximation or simpler estimates (see below). 

The work of Fryer and Heger~\cite{fryer-00} is an improvement over past collapse investigations because it starts with rotating progenitors evolved to collapse with a stellar evolution code (which incorporates angular momentum transport via an approximate diffusion scheme)~\cite{heger-98}, incorporates realistic equations of state and neutrino transport, and follows the collapse to late times. 
The values of total angular momentum of the inner cores of Fryer and Heger ($0.95$--$1.9\times 10^{49}\,{\rm g\,cm^2\,s^{-1}}$) are lower than has often been assumed in studies of the GW emission from core collapse. Note that the total specific angular momentum of these core models may be lower by about a factor of 10 if magnetic fields were included in the evolution of the progenitors~\cite{akiyama-02, heger-02, heger-02a}.

FHH's~\cite{fryer-02} numerical quadrupole estimate of the GWs from polar oscillations in the collapse simulations of Fryer and Heger~\cite{fryer-00} predicts a peak dimensionless amplitude $h_{pk}=4.1\times 10^{-23}$ (for $d$=$10\,{\rm Mpc}$), emitted at $f_{GW} \approx 20\,{\rm Hz}$.  The radiated energy $E_{GW} \sim 2 \times 10^{44} {\rm ergs}$. This signal would be just out of the detectability range of the LIGO--II detector.

The cores in the simulations of Fryer and Heger~\cite{fryer-00} are not compact enough (or rotating rapidly enough) to develop bar instabilities during the collapse and initial bounce phases.  However, the explosion phase ejects a good deal of low angular momentum material along the poles in their evolutions.  Therefore, about $1\,{\rm s}$ after the collapse, $\beta$ becomes high enough in their models to exceed the secular bar instability limit.  The $\beta$ of their model with the least angular momentum actually exceeds the dynamical bar instability limit as well (it contracts to a smaller radius and thus has a higher spin rate then the model with higher angular momentum).  FHH (and~\cite{fryer-02b}) compute an upper limit (via equation~[\ref{equation:hbar}]) to the emitted amplitude from their dynamically unstable model of $h \sim 3\times 10^{-22}$ (if coherent emission from a bar located at $10\, {\rm Mpc}$ persists for 100 cycles).  The corresponding frequency and maximum power are $f_{GW} \approx 10^{3}\,{\rm Hz}$ and $P_{GW} = 10^{53} {\rm erg\,s^{-1}}$.  LIGO--II should be able to detect such a signal (see figure~\ref{figure:ligo}, where FHH's upper limit to $h$ for this dynamical bar--mode is identified).

As mentioned above, the proto--neutron stars of Fryer and Heger are likely unstable to the development of secular bar instabilities.
The GW emission from proto--neutron stars that are secularly unstable to the bar--mode has been examined by Lai and Shapiro~\cite{lai-95, lai-01}.  Because the timescale for secular evolution is so long, 3d hydrodynamics simulations of the nonlinear development of a secular bar can be impractical. To bypass this difficulty, Lai~\cite{lai-01} considers only incompressible fluids, for which there are exact solutions for (Dedekind and Jacobi--like) bar development.  He predicts that such a bar located at $10 {\rm Mpc}$ would emit GWs with a peak characteristic amplitude $h\sim 10^{-21}$, if the bar persists for $10^2$--$10^4$ cycles. The maximum $f_{GW}$ of the emitted radiation is in the range $10^2-10^3 {\rm Hz}$.  This type of signal should be easily detected by LIGO--I (although detection may require a technique like the fast chirp transform method of Jenet and Prince~\cite{jenet-00} due to the complicated phase evolution of the emission).

FHH predict that a fragmentation instability is unlikely to develop during core-collapse SNe because the cores have central density maxima (see also~\cite{fryer-02b}).  However, they do give estimates (calculated via equations~[\ref{equation:hbin}, \ref{equation:pbin}]) for the amplitude, power, and frequency of the emission from such an instability: $h_{pk}\sim 2 \times 10^{-22}$, $P_{GW} = 10^{54} {\rm erg\,s^{-1}}$, $f_{GW} \approx 2 \times 10^{3}\,{\rm Hz}$. Again, this signal would fall just beyond the upper limit of LIGO--II's frequency range.   Drawing on analogy with fragmentation observed in star formation simulations, Davies et al.~\cite{davies-02} have recently suggested that the fragmentation of collapsing massive stellar cores and the subsequent gravitational radiation driven coalescence of the fragments, may power gamma-ray bursts (see also~\cite{kobayashi-02}).  They say that due to kick velocity constraints, this would be possible only for supernovae that leave behind black hole remnants with $M \gtrsim 12 M_{\odot}$ (formed via the merger of the fragments).  Under the assumption that the collapsing core splits into two fragments (each with $M = 0.7 M_{\odot}$), they estimate that LIGO--II should detect the merger of $\sim 400$ core collapse fragments per year.  Note that the signal strength would be even greater if the total mass of the black hole remnant were indeed $12 M_{\odot}$ (however the formation rate for these more massive objects is likely lower than that of lower mass cores).

The GW emission from $r$--mode unstable neutron star remnants of core collapse SNe would be easily  detectable if $\alpha_{max} \sim 1$ (which is likely not physical; see section~\ref{section:aic-mech}).  Multiple GW bursts will occur as material falls back onto the neutron star and results in repeat episodes of $r$--mode growth (note that a single $r$--mode episode can have multiple amplitude peaks~\cite{lindblom-01}).  FHH calculate that the characteristic amplitude of the GW emission from this $r$--mode evolution tracks from $6$--$1 \times 10^{-22}$, over a frequency range of $10^3$--$10^2\,{\rm Hz}$ (see section~[\ref{section:aic-num}] for details). They estimate the emitted energy to exceed $10^{52}\,{\rm ergs}$.

If the collapse remnant is a black hole, GWs will be emitted as the accreting black hole "rings down." Even with very optimistic accretion scenarios, FHH conclude that such radiation will be of very low amplitude and beyond the upper frequency reach of LIGO--II (see~\cite{fryer-02} for details).

General relativity has been more fully accounted for in the core collapse studies of Dimmelmeier, Font, and M\"{u}ller~\cite{dimmelmeier-01, dimmelmeier-02a, dimmelmeier-02b}, which build on the Newtonian, axisymmetric collapse simulations of Zwerger and M\"{u}ller~\cite{zwerger-97}.  
In all, they have followed the collapse evolution of 26 different models, with both Newtonian and general relativistic simulations. As in the work of Zwerger and M\"{u}ller, the different models are characterized by varying degrees of differential rotation, initial rotation rates, and adiabatic indices. They use the conformally flat metric to approximate the space time geometry~\cite{cook-96} in their relativistic hydrodynamics simulations.  This approximation gives the exact solution to Einstein's equations in the case of spherical symmetry.  Thus, as long as the collapse is not significantly aspherical, the approximation is relatively accurate.  However, the conformally flat condition does eliminate GW emission from the spacetime.  Because of this, Dimmelmeier, Font, and M\"{u}ller used the quadrupole approximation to compute the characteristics of the emitted GW signal (see~\cite{zwerger-97} for details).

\begin{figure}[h]
 \def\epsfsize#1#2{0.4#1}
 \centerline{\epsfbox{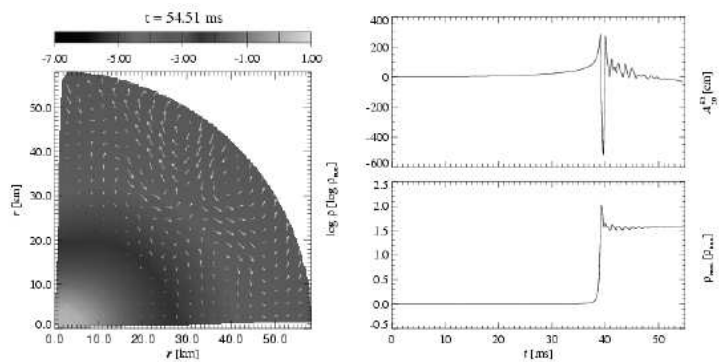}}
  \caption{\it Movies from the evolution of the regular collapse model, A3B2G4, of Dimmelmeier et al.~\cite{dimmelmeier-02b}.  The left frame contains the 2d evolution of the logarithmic density.  The upper and lower right frames display the evolutions of the gravitational wave amplitude and the maximum density, respectively.}
 \label{figure:dfm-mov1}
\end{figure}

\begin{figure}[h]
 \def\epsfsize#1#2{0.4#1}
 \centerline{\epsfbox{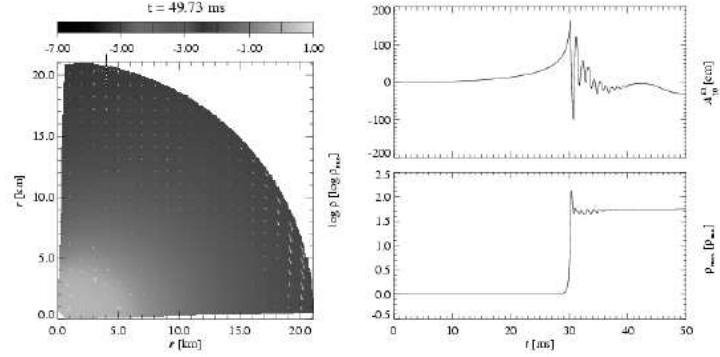}}
  \caption{\it Same as figure~\ref{figure:dfm-mov1}, but for rapid collapse model A3B2G5 of Dimmelmeier et al.~\cite{dimmelmeier-02b}.  }
 \label{figure:dfm-mov2}
\end{figure}

\begin{figure}[h]
 \def\epsfsize#1#2{0.4#1}
 \centerline{\epsfbox{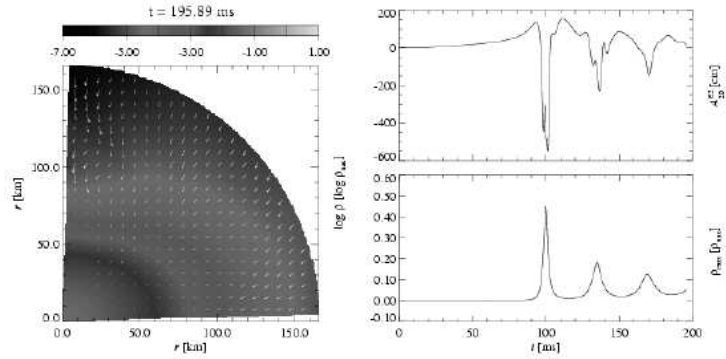}}
  \caption{\it Same as figure~\ref{figure:dfm-mov1}, but for multiple collapse model A2B4G1 of Dimmelmeier et al.~\cite{dimmelmeier-02b}.  }
 \label{figure:dfm-mov3}
\end{figure}

\begin{figure}[h]
 \def\epsfsize#1#2{0.4#1}
 \centerline{\epsfbox{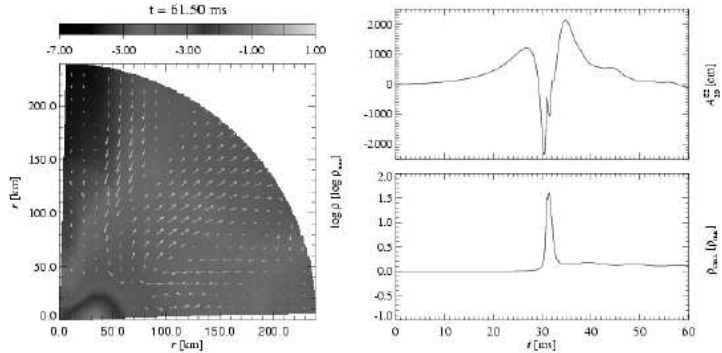}}
  \caption{\it Same as figure~\ref{figure:dfm-mov1}, but for rapid, differentially rotating collapse model A4B5G5 of Dimmelmeier et al.~\cite{dimmelmeier-02b}.  }
 \label{figure:dfm-mov4}
\end{figure}

The general relativistic simulations of Dimmelmeier et al.\ showed the three different types of collapse evolution (and corresponding gravitational radiation signal) seen in the Newtonian simulations of Zwerger and M\"{u}ller (regular collapse--Type I signal; multiple bounce collapse--Type II signal; and rapid collapse--Type III signal).  However, relativistic effects sometimes led to a different collapse type than in the Newtonian case.  This is because general relativity did indeed counteract the stabilizing effects of rotation and led to much higher bounce densities (up to $700\%$ higher).  They found that multiple bounce collapse is much rarer in general relativistic simulations (occurring in only two of their models).  When multiple bounce does occur, relativistic effects shorten the time interval between bounces by up to a factor of four.  Movies of the simulations of four models from Dimmelmeier et al.~\cite{dimmelmeier-02b} are shown in figures~\ref{figure:dfm-mov1}--\ref{figure:dfm-mov4}.  The four evolutions shown include a regular collapse (~\ref{figure:dfm-mov1}), a rapid collapse (~\ref{figure:dfm-mov2}), a multiple bounce collapse (~\ref{figure:dfm-mov3}), and a very rapidly and differentially rotating collapse (~\ref{figure:dfm-mov4}).   The left frames of each movie contain the 2d evolution of the logarithmic density.  The upper and lower right frames display the evolutions of the gravitational wave amplitude and the maximum density, respectively.  These movies can also be viewed at~\cite{dfm-mov}. 

\clearpage

Dimmelmeier et al.\ found that models for which the collapse type was the same in both Newtonian and relativistic simulations, had lower GW amplitudes $h_{pk}$ in the relativistic case.  This is because the Newtonian models were less compact at bounce and thus had material with higher densities and velocities at larger radii.  Both higher and lower values of $h_{pk}$ were observed in models for which the collapse type changed. Overall, the range of $h_{pk}$ ($4 \times 10^{-24}$--$3 \times 10^{-23}$, for a source located at $10\,{\rm Mpc}$) seen in the relativistic simulations was quite close to the corresponding Newtonian range.  The average $E_{GW}$ was somewhat higher in the relativistic case ($1.5 \times 10^{47}\,{\rm ergs}$ compared to the Newtonian value of $6.4 \times 10^{46}\,{\rm ergs}$).  The overall range of GW frequencies observed in their relativistic simulations ($60$--$1000\,{\rm Hz}$) was close to the Newtonian range.  They did note that relativistic effects always caused the characteristic frequency of emission, $f_{GW}$, to increase (up to five--fold).  For most of their models, this increase in $f_{GW}$ was not accompanied by an increase in $h_{pk}$.  This means that relativistic effects could decrease the detectability of GW signals from some core collapses. However, the GW emission from the models of Dimmelmeier et al.\ could be detected by the first generation of ground-based interferometric detectors if the sources were fortuitously located in the Local Group of galaxies.  A catalog containing the signals and spectra of the GW emission from all of their models can be found at~\cite{dfm-gw}.

Fully general relativistic collapse simulations (i.e., without the conformally flat approximation) have been performed by Shibata~\cite{shibata-00ptp}.  He used an axisymmetric code that solves the Einstein equations in Cartesian coordinates and the hydrodynamics equations in cylindrical coordinates.  The use of the Cartesian grid eliminates the presence of singularities and allows for stable, long--duration axisymmetric simulations~\cite{alcubierre-99}.  The focus of this work was the effect of rotation on the criteria for prompt black hole formation.  Shibata found that if the parameter $q=J/M^2$ is less than $0.5$, black hole formation occurred for rest masses slightly greater than the maximum mass of spherical stars.   However, for $0.5<q<1$, the maximum stable rest mass is increased by $\sim 70$--$80\%$.  The results are only weakly dependent on the initial rotation profile.  Shibata did not compute the GW emission in his collapse simulations.

The new general relativistic hydrodynamics simulations of Zanotti, Rezzolla, and Font~\cite{zanotti-02} suggest that a torus of neutron star matter surrounding a black hole remnant may be a stronger source of GWs than the collapse itself.  They used a high resolution shock-capturing hydrodynamics method in conjunction with a static (Schwarzschild) spacetime to follow the evolution of ``toroidal neutron stars.''  Their results indicate that if a toroidal neutron star (with constant specific angular momentum) is perturbed, it could undergo regular oscillations.  They estimate that the resulting GW emission would have a characteristic amplitude $h_c$ ranging from $6 \times 10^{-24}$--$5 \times 10^{-23}$, for ratios of torus mass to black hole mass in the range $0.1$--$0.5$.  (These amplitude values are likely underestimated because the simulations of Zanotti et al.\ are axisymmetric.)  The corresponding frequency of emission is $f_{GW}\sim 200\,{\rm Hz}$.  The values of $h_c$ and $f_{GW}$ quoted here are for a source located at $10\,{\rm Mpc}$.  This emission would be just outside the range of LIGO-II (see figure~\ref{figure:ligo}). Further numerical investigations, which study tori with non-constant angular momenta and include the effects of self-gravity and black hole rotation, are needed to confirm these predictions. Movies from the simulations of Zanotti et al.\ can be viewed at~\cite{zrf-mov}.

Magnetized tori around rapidly spinning black holes (formed via either core collapse or neutron star--black hole coalescence) have recently been examined in the theoretical study of van Putten and Levinson~\cite{putten-02}. They find that such a torus--black hole system can exist in a suspended state of accretion if the ratio of poloidal magnetic field energy to kinetic energy $E_B/E_k$ is less than $0.1$.  They estimate that $\sim 10\%$ of the spin energy of the black hole will be converted to gravitational radiation energy through multipole mass moment instabilities that develop in the torus.  If a magnetized torus-black hole system located at $10\,{\rm Mpc}$ is observed for $2 \times 10^4$ rotation periods, the characteristic amplitude of the GW emission is $\sim 6 \times 10^{-20}$.  It is possible that this emission could take place at several frequencies.  Observations of x--ray lines from gamma--ray bursts (which are possibly produced by these types of systems) could constrain these frequencies by providing information regarding the angular velocities of the tori (preliminary estimates from observations suggest $f_{GW} \sim 500\,{\rm Hz}$, placing the radiation into a range detectable by LIGO-I).  

\subsubsection{Simulations of Convective Instabilities}

Convectively driven inhomogeneities in the density distribution of the outer regions of the nascent neutron star and anisotropic neutrino emission are other sources of GW emission during the collapse/explosion~\cite{burrows-96, mueller-98}.  GW emission from these processes results from small-scale asphericities, unlike the large--scale motions responsible for GW emission from aspherical collapse and non--axisymmetric global instabilities.  Note that Rayleigh-Taylor instabilities also induce time--dependent quadrupole moments at composition interfaces in the stellar envelope.  However, the resultant GW emission is too weak to be detected because the Rayleigh-Taylor instabilities occur at very large radii~\cite{mueller-98}.

Burrows and Hayes~\cite{burrows-96} have suggested that the large proper motions observed in pulsars may result from asymmetrical core collapse SNe.  They have computed the resultant GW emission under the assumption that such asymmetries were caused by convection and anisotropic neutrino emission during the collapse and explosion.  Hydrodynamic simulations and theoretical investigations suggest that asymmetries present in the star prior to collapse (in part due to convection during silicon and oxygen burning) will be amplified during the collapse~\cite{bazan-94, lai-00}.  Based on these arguments, Burrows and Hayes induced a mass dipole anisotropy of $< 0.1\%$ in the pre-collapse core used in their ``exploratory'' collapse simulation.  The initial quadrupole anisotropy present in their model was $0.012$.  They imposed no aspherical perturbations on the initial velocities.  The collapse was followed in 1D inside a radius of $15\,{\rm km}$ and in 2D (with azimuthal symmetry) to larger radii.  To reduce the computational burden of this exploratory simulation, they artificially hardened the neutrino spectrum to drive the explosion at a faster rate (note that this practice was observed to minimize the GW emission in the simulations of M\"{u}ller and Janka~\cite{mueller-97}, see below).

\begin{figure}[h]
 \def\epsfsize#1#2{0.25#1}
 \centerline{\epsfbox{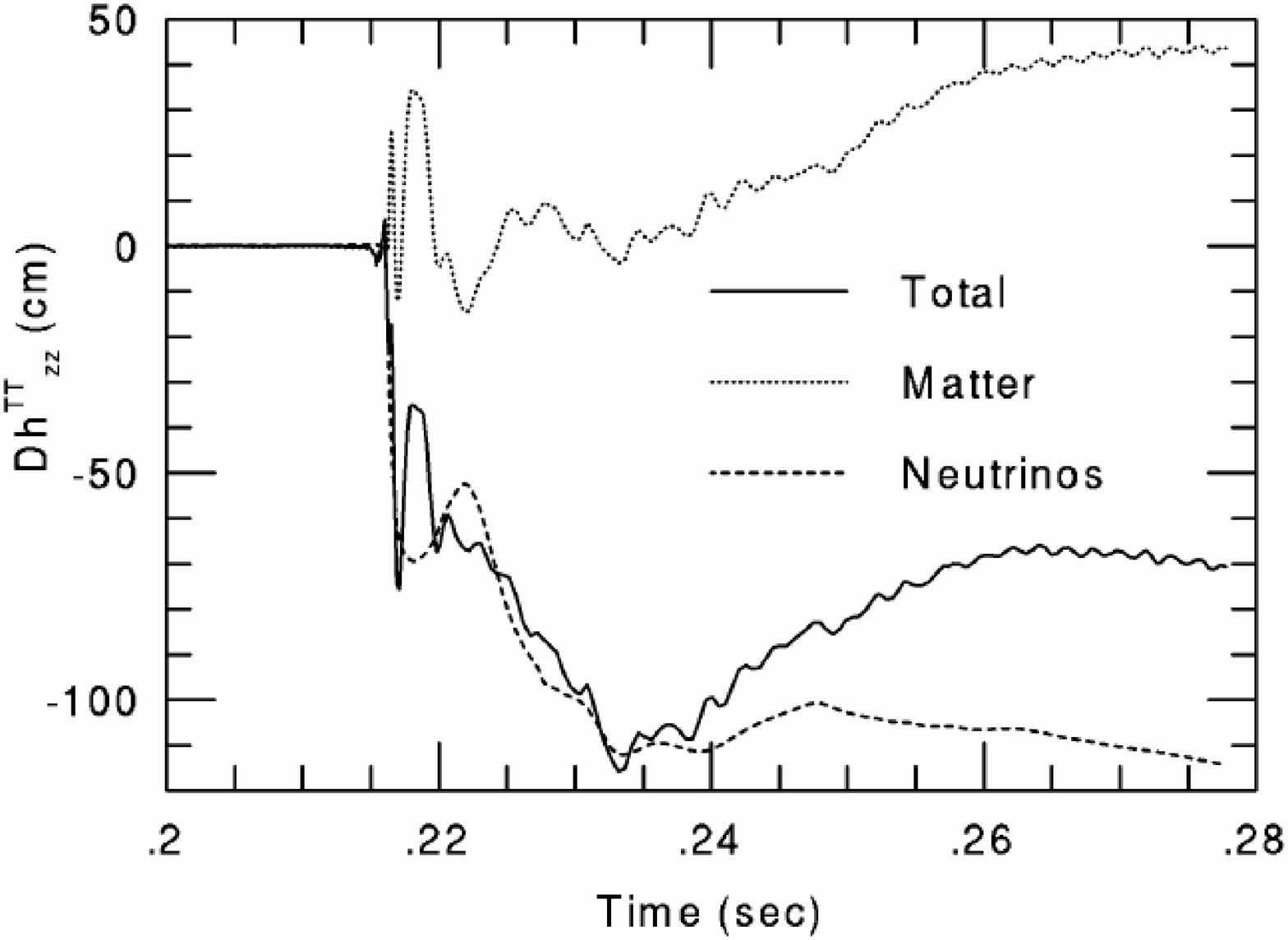}}
 \caption{\it The gravitational waveform (including separate matter and neutrino contributions) from the collapse simulations of Burrows and Hayes~\cite{burrows-96}.  The curves plot the product of the gravitational wave amplitude and distance to the source as a function of time. (Figure 3 of~\cite{burrows-96}; used with permission.)}
 \label{figure:burrows}
\end{figure}

The initial asymmetries were observed to grow in Burrows and Hayes' calculation.  The nascent neutron star developed a recoil speed of $\sim 530\,{\rm km\,s^{-1}}$.  Aspherical mass motion was responsible for the vast majority of the recoil.  Only 16\% of the recoil velocity could be attributed to anisotropic neutrino emission.  The gravitational waveform from this simulation (including separate matter and neutrino contributions) is show in Figure~\ref{figure:burrows}. The contributions of matter motion and neutrino emission to the GW amplitude $h$ were of opposite sign.  Anisotropic neutrino radiation was basically the sole contributor to $h$ for the first $20\,{\rm ms}$, after which the contribution from mass motion became significant.  The peak amplitude calculated was $h_{pk}\sim 3 \times 10^{-24}$, for a source located at $10\,{\rm Mpc}$. Most ($\sim 70\%$) of the emitted $E_{GW}\sim 2.0\times 10^{45}\,{\rm ergs}$  came from mass motion (because the mass quadrupole was changing more rapidly as the bounce occurred).  The frequency of the GWs emitted was in the range $10 < f_{GW} < 100\,{\rm Hz}$.  Burrows and Hayes predicted LIGO--II should be able to detect this emission, with a signal-to-noise ratio of 10, from a core collapse that fortuitously happened to occur in our Galaxy (with $d$=$10\,{\rm kpc}$).

The study of Nazin and Postnov~\cite{nazin-97} predicts a lower limit for $E_{GW}$ emitted during an asymmetric core collapse SNe (where such asymmetries could be induced by both aspherical mass motion and neutrino emission).  They assume that observed pulsar kicks are solely due to asymmetric collapse.  They suggest that the energy associated with the kick ($M v^2/2$, where $M$ and $v$ are the mass and velocity of the neutron star) can be set as a lower limit for $E_{GW}$ (which can be computed without having to know the mechanism behind the asymmetric collapse).  From observed pulsar proper motions, they estimate the degree of asymmetry $\epsilon$ present in the collapse and the corresponding characteristic GW amplitude ($h \propto \sqrt{\epsilon}$).  This amplitude is $3 \times 10^{-25}$ for a source located at $10\,{\rm Mpc}$ and emitting at $f_{GW}=1\,{\rm kHz}$.

M\"{u}ller and Janka performed both 2D and 3D simulations of convective instabilities in the proto--neutron star and hot bubble regions during the first second of the explosion phase of a Type II SNe~\cite{mueller-97}.  They numerically computed the GW emission from the convection induced aspherical mass motion and neutrino emission in the quadrupole approximation (for details, see section 3 of their paper).

\begin{figure}[h]
 \def\epsfsize#1#2{0.4#1}
 \centerline{\epsfbox{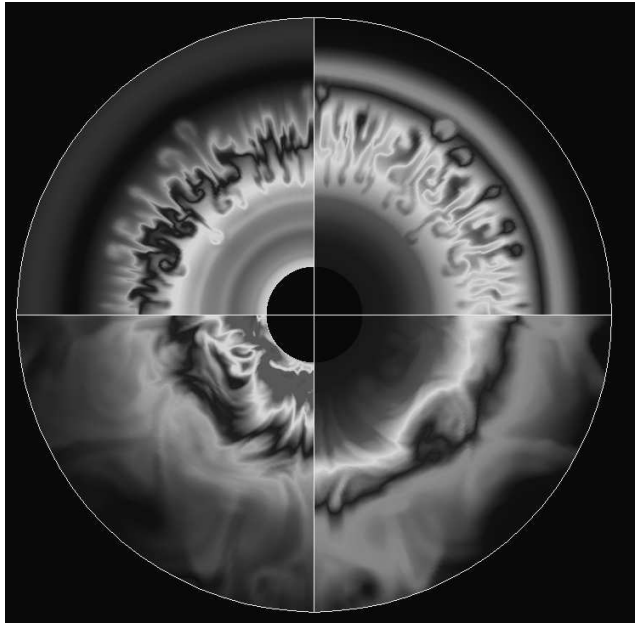}}
 \caption{\it Convective instabilities inside the proto--neutron star in the 2D simulation of M\"{u}ller and Janka~\cite{mueller-97}.  The evolutions of the temperature (left panels) and logarithmic density (right panels) distributions are shown for the radial region $15$--$95\,{\rm km}$.  The upper and lower panels correspond to times $12$ and $21\,{\rm ms}$, respectively, after the start of the simulation.  The temperature values range from $2.5\times 10^{10}$ to $1.8\times 10^{11}\,{\rm K}$.  The values of the logarithm of the density range from $10.5$ to $13.3\,{\rm g\,cm^{-3}}$.  The temperature and density both increase as the colors change from blue to green, yellow, and red.  (Figure 7 of~\cite{mueller-97}; used with permission.)}
 \label{figure:rgb}
\end{figure}

For typical iron core masses, the convectively unstable region in the proto--neutron star extends over the inner $0.7$--$1.20 M_{\odot}$ of the core mass (this corresponds to a radial range of $\sim 10$--$50\,{\rm km}$). The convection in this region, which begins approximately $10$--$20\,{\rm ms}$ after the shock forms and may last for $\approx 20\,{\rm ms}$--$1\,{\rm s}$, is caused by unstable gradients in entropy and/or lepton number resulting from the stalling of the prompt shock and deleptonization outside the neutrinosphere.  M\"{u}ller and Janka's simulations of convection in this region began with the 1D, non-rotating, $12\,{\rm ms}$ post--bounce model of Hillebrandt\cite{hillebrandt-87}.  This model included general relativistic corrections that had to be relaxed away prior to the start of the Newtonian simulations.  Neutrino transport was neglected in these runs (see section 2.1 of~\cite{mueller-97} for justification); however, a sophisticated equation of state was utilized. Figure~\ref{figure:rgb} shows the evolution of the temperature and density distributions in the 2D simulation of M\"{u}ller and Janka.  

\begin{figure}[h]
 \def\epsfsize#1#2{0.5#1}
 \centerline{\epsfbox{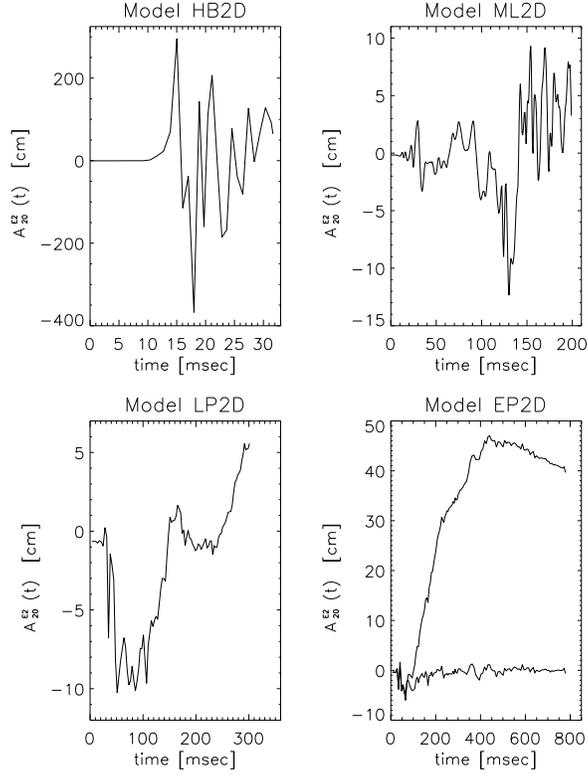}}
 \caption{\it Quadrupole amplitudes $A^{E2}_{20}[cm]$ from convective instabilities in various models of ~\cite{mueller-97}. The upper left panel is the amplitude from a 2D simulation of proto--neutron star convection.  The other three panels are amplitudes from 2D simulations of hot bubble convection.  The imposed neutrino flux in the hot bubble simulations increases from the top right model through the bottom right model.  (Figure 18 of~\cite{mueller-97}; used with permission.)}
 \label{figure:convection}
\end{figure}

The peak GW amplitude resulting from convective mass motions in these simulations of the proto--neutron star was $\approx 3 \times 10^{-24}$ in 2D and $\approx 2 \times 10^{-25}$ in 3D, for $d$=$10\,{\rm Mpc}$.  The emitted energy was $9.8\times 10^{44}\,{\rm ergs}$ in 2D and $1.3\times 10^{42}\,{\rm ergs}$ in 3D. The power spectrum peaked at frequencies of $200$--$600\,{\rm Hz}$ in 2D and $100$--$200\,{\rm Hz}$ in 3D.  Such signals would not be detectable with LIGO--II. The reasons for the differences between the 2D and 3D results include smaller convective elements and less under-- and overshooting in 3D.  The relatively low angular resolution of the 3D simulations may have also played a role.  The quadrupole gravitational wave amplitude $A^{E2}_{20}$ from the 2D simulation is shown in the upper left panel of figure~\ref{figure:convection} (see~\cite{zwerger-97, thorne-80} for expressions relating $A^{E2}_{20}$ to $h$).

Convection in the hot bubble region between the shock and neutrinosphere arises because of an unstable entropy gradient resulting from neutrino heating.  This unstable region extends over the inner mass range $1.25$--$1.40 M_{\odot}$ (corresponding to a radial range of $\approx 100$--$1000\,{\rm km}$).  Convection in the hot bubble begins $\approx 50$--$80\,{\rm ms}$ after shock formation and lasts from $\approx 100$--$500\,{\rm ms}$.  Only 2D simulations were performed in this case.  These runs started with a $25\,{\rm ms}$ post--bounce model provided by Bruenn.  A simple neutrino transport scheme was used in the runs and an imposed neutrino flux was located inside the neutrinosphere.  Due to computational constraints, the computational domain did not include the entire convectively unstable region inside the proto--neutron star (thus this set of simulations only accurately models the convection in the hot bubble region, not in the proto--neutron star).

The peak GW amplitude resulting from these 2D simulations of convective mass motions in the hot bubble region was $h_{pk}\approx 5 \times 10^{-25}$, for $d$=$10\,{\rm Mpc}$.  The emitted energy was $\lesssim  2\times 10^{42}\,{\rm ergs} $. The energy spectrum peaked at frequencies of $50$--$200\,{\rm Hz}$.  As the explosion energy was increased (by increasing the imposed neutrino flux), the violent convective motions turn into simple rapid expansion.  The resultant frequencies drop to $f_{GW}\sim 10\,{\rm Hz}$.  The amplitude of such a signal would be too low to be detectable with LIGO--II.

M\"{u}ller and Janka estimated the GW emission from the convection induced anisotropic neutrino radiation in their simulations (see~\cite{mueller-97} for details).  They could not directly compute the characteristics of the GW emission from neutrinos because their proto--neutron star simulations did not include neutrino transport and the neutrino transport in their hot bubble simulations was only 1D.  They found that the amplitude of the GWs emitted can be a factor of $5$--$10$ higher than the GW amplitudes resulting from convective mass motion.  However, their results indicate that the $E_{GW}$ coming from anisotropic neutrino emission is only a small fraction of that emitted by convective mass motion.  

Note that the qualitative results (including $h$ and $E_{GW}$ magnitudes) of Burrows and Hayes'~\cite{burrows-96} ``exploratory'' 2D calculation are in rough agreement with the more detailed 2D simulations of M\"{u}ller and Janka .  However, both investigations have shortcomings.  Necessary improvements would include more realistic initial models (which include rotation and are produced with multi-dimensional collapse simulations to better estimate initial asphericities), use of sophisticated neutrino transport schemes, inclusion of general relativistic effects, and higher resolution 3D runs.

\clearpage

\subsection{Going Further}
\label{section:sne-gf}

The background of GW emission from a population of core-collapse SNe at cosmological distances may be detectable by LIGO--II, according to Ferrari, Matarrese, and Schneider~\cite{ferrari-99}.
They determined the SNe rate as a function of redshift using observations to determine the evolution of the star formation rate.  Only collapses that lead to black hole formation were considered.  This simplified the study because the GW emission from such collapses is generally a function of just the black hole mass and angular momentum. They found that the stochastic background from these sources is not continuous and suggest that this could be used to optimize detection strategies.  The maximum GW spectral strain amplitude they computed was in the range $10^{-28}$--$10^{-27}\,{\rm Hz}$, at frequencies of a few times $10^2\,{\rm Hz}$.  Such a signal may be detected by a pair of LIGO--II detectors.

\clearpage
\newpage


\section{Collapse of Population III Stars}
\label{section:pop3}

\subsection{Collapse Scenario}
\label{subsection:pop3-scen}

The first generation of stars to form in the early universe are known as Population III stars (formed at redshifts $z \gtrsim 5$).  Theoretical and computational evidence suggests that Population III stars may have had masses $\gtrsim 100 M_{\odot}$~\cite{abel-02, abel-00, fryer-01}.  Because these massive stars contained no metals, it was possible for them to form directly and then evolve with very low stellar winds and thus very little mass loss.  If the mass of a nonrotating Population III star is $\gtrsim 260 M_{\odot}$, its fate is to collapse directly to a black hole at the end of its life~\cite{fryer-01}.  If rotational support prevents the star from direct collapse to a black hole, explosive thermonuclear burning will cause the star to undergo a giant hypernova explosion.  Prior to black hole formation, the rotating, collapsed core will have a mass of $50$--$70 M_{\odot}$ and a radius of $1000$--$2000\,{\rm km}$.  Note that because these massive stars evolve so quickly (in a few million years~\cite{baraffe-01}), the events associated with their deaths will take place at roughly the redshifts of their births.

\subsection{Formation Rate}
\label{section:pop3-form}

The formation rate of Population III stars can be indirectly estimated from the re-ionization fraction of the early universe, which was re-ionized by light from these stars~\cite{hogan-79, carr-84}.  Using estimates of the ultraviolet light emission from Population III stars, their ionization efficiency, and the re-ionization fraction of the early universe, one can determine that about $0.01\%$--$1\%$ of the universe's baryonic matter was found in these very massive stars.  This corresponds to $\sim 10^4$--$10^7$ Population III stars in a $10^{11} M_{\odot}$ galaxy and thus a collapse rate that is $\lesssim 10^{-3}\,{\rm yr^{-1}}$.  Thus, a reasonable occurrence rate can be found for an observation (luminosity) distance of $\sim 50\,{\rm Gpc}$ (which corresponds to a redshift of $z$=$5$, in the cosmology used by ~\cite{fryer-02}).  Uncertainties in the assumptions used make this formation rate uncertain to within a few orders of magnitude.~\cite{abel-00, fryer-01, abel-02, fryer-02}  It is unknown how many Population III stars will collapse directly to black holes.  

\subsection{GW Emission Mechanisms}
\label{section:pop3-mech}

The GW emission mechanisms related to the collapse of Population III stars are a subset of those discussed in the sections on AIC and SNe/collapsars.  These include aspherical collapse, global rotational and fragmentation instabilities that may arise during the collapse/explosion and in the collpase remnant (prior to black hole formation), and the ``ring--down'' of the remnant black hole.

\subsection{Numerical Predictions of GW Emission}
\label{section:pop3-num}

The GW emission from the collapse of Population III stars has recently been investigated by Fryer and collaborators  (Fryer, Woosley, and Heger~\cite{fryer-01}, FHH~\cite{fryer-02}, and Fryer, Holz, Hughes, and Warren~\cite{fryer-02b}).  The collapse simulations of Fryer, Woosley, and Heger again started with rotating collapse progenitors that had been evolved with a stellar evolution code~\cite{heger-98}.  The initial models used by the evolution code were in rigid rotation with a surface ratio of centrifugal to gravitational forces of $20\%$ (this ratio is seen in current observations of O stars).

The results of Fryer, Woosley, and Heger suggest that the collapse remnant (prior to black hole formation) is susceptible to the development of a secular bar--mode instability.  However, at $z>5$, the GW emission would be 
redshifted out of LIGO--II's frequency range.  At $z$=$5$, $h_{pk}=8\times 10^{-23}$, with a corresponding frequency of $10\,{\rm Hz}$~\cite{fryer-02, fryer-02b}.  Even if such a signal persists for a hundred cycles, it would likely be undetectable by LIGO--II.  Note that these signal strengths are orders of magnitude lower than the qualitative estimates of signal strength given in Carr, Bond, and Arnett~\cite{carr-84}.

LIGO--II may be able to detect the GW emission from binary clumps formed via a fragmentation instability. If such a signal is emitted at $z$=$5$ and persists for 10 cycles, $h$ would be $\sim 10^{-22}$, over a frequency range of $10$--$100\,{\rm Hz}$~\cite{fryer-02, fryer-02b}.  The likelihood of the development of a fragmentation instability is diminished by the fact that the off-center density maxima present in the simulations of Fryer, Woosley, and Heger are not very pronounced~\cite{fryer-02p}.  

The ``ring--down'' of the black hole remnant will likely be strong because Fryer, Woosley, and Heger observe a high accretion rate after collapse.  FHH estimate that for a source located at $z$=$20$, the GWs would be redshifted out of LIGO--II's bandwidth.  However, for a source at $z$=$5$, $h_{pk}\sim 6 \times 10^{-23}$ and the frequency range is $20$--$70\,{\rm Hz}$.  This signal may be marginally detectable with LIGO-II (see figure~\ref{figure:ligo}).

\newpage


\section{Collapse of Supermassive Stars}
\label{section:sms}

\subsection{Collapse Scenario}
\label{subsection:sms-scen}

There is a large body of observational evidence that supermassive black holes (SMBHs, $M \gtrsim 10^6 M_{\odot}$) exist in the centers of many, if not most galaxies (see, e.g., the reviews of Rees~\cite{rees-98} and Macchetto~\cite{macchetto-99}).  The masses of SMBHs in the centers of more than 45 galaxies have been estimated from observations~\cite{ferrarese-00} and there are more than 30 galaxies in which the presence of a SMBH has been confirmed~\cite{kormendy-01}.

One of the possible formation mechanisms for SMBHs involves the gravitational collapse of supermassive stars (SMSs). The timescale for this formation channel is short enough to account for the presence of SMBHs at redshifts $z>6$~\cite{janka-02}. Supermassive stars may contract directly out of the primordial gas, if radiation and/or magnetic field pressure prevent fragmentation~\cite{haehnelt-93, eisenstein-95, haehnelt-98, loeb-94, bromm-99, abel-00}.  Alternatively, they may build up from fragments of stellar collisions in clusters~\cite{sanders-70, begelman-78}.  Supermassive stars are radiation dominated, isentropic and convective~\cite{shapiro-83, zeldovich-71, loeb-94}. Thus, they are well represented by an $n$=$3$ polytrope.  If the star's mass exceeds $10^6 M_{\odot}$, nuclear burning and electron/positron annihilation are not important.

After formation, an SMS will evolve through a phase of quasistationary cooling and contraction.  If the SMS is rotating when it forms, conservation of angular momentum requires that it spins up as it contracts.  There are two possible evolutionary regimes for a cooling SMS.  The path taken by an SMS depends on the strength of its viscosity and magnetic fields and on the nature of its angular momentum distribution.

In the first regime, viscosity or magnetic fields are strong enough to enforce {\it uniform} rotation throughout the star as it contracts.  Baumgarte and Shapiro~\cite{baumgarte-99} have studied the evolution of a uniformly rotating SMS up to the onset of relativistic instability.  They demonstrated that a uniformly rotating, cooling SMS will eventually spin up to its mass shedding limit.  The mass shedding limit is encountered when matter at the star's equator rotates with the Keplerian velocity.  The limit can be represented as $\beta_{shed}=(T/|W|)_{shed}$.  In this case, $\beta_{shed}=9\times 10^{-3}$. The star will then evolve along a mass shedding sequence, losing both mass and angular momentum.  It will eventually contract to the onset of relativistic instability~\cite{iben-63, chandrasekhar-64a, chandrasekhar-64b, shapiro-83, janka-02}.

Baumgarte and Shapiro used both a second-order, post-Newtonian approximation and a fully general relativistic numerical code to determine that the onset of relativistic instability occurs at a ratio of $R/M \sim 450$, where $R$ is the star's radius and $G$=$c$=$1$ in the remainder of this section.  Note that a second-order, post-Newtonian approximation was needed because rotation stabilizes the destabilizing role of nonlinear gravity at the first post-Newtonian level.
If the mass of the star exceeds $10^6 M_{\odot}$, the star will then collapse and possibly form a SMBH. If the star is less massive, nuclear reactions may lead to explosion instead of collapse.

The major result of Baumgarte and Shapiro's work is that the universal values of the following ratios exist for the critical configuration at the onset of relativistic instability:  $T/|W|$, $R/M$, and $J/M^2$.  These ratios are completely independent of the mass of the star or its prior evolution.  Because uniformly rotating SMSs will begin to collapse from a universal configuration, the subsequent collapse and the resulting gravitational waveform will be unique.

In the opposite evolutionary regime, neither viscosity nor magnetic fields are strong enough to enforce uniform rotation throughout the cooling SMS as it contracts.  In this case it has been shown that the angular momentum distribution is conserved on cylinders during contraction~\cite{bodenheimer-73}.  Because viscosity and magnetic fields are weak, there is no means of redistributing angular momentum in the star.  So even if the star starts out rotating uniformly, it cannot remain so.

The star will then rotate {\it differentially} as it cools and contracts.  In this case, the subsequent evolution depends on the star's initial angular momentum distribution, which is largely unknown.  One possible outcome is that the star will spin up to mass shedding (at a different value of $\beta_{shed}$ than a uniformly rotating star) and then follow an evolutionary path that may be similar to that described by Baumgarte and Shapiro~\cite{baumgarte-99}.  The alternative outcome is that the star will encounter the dynamical bar instability prior to reaching the mass shedding limit.  New and Shapiro~\cite{new-01a, new-01b} have demonstrated that a bar--mode phase is likely to be encountered by differentially rotating SMSs with a wide range of initial angular momentum distributions.  This mode will transport mass and angular momentum outward and thus may hasten the onset of collapse.

\subsection{Formation Rate}
\label{section:sms-form}

An estimate of the rate of the collapse of SMSs can be derived from the quasar luminosity function.  Haehnelt \cite{haehnelt-94} has used the quasar luminosity function to compute the rate of GW bursts from supermassive black holes, assuming that each quasar emits one such burst during its lifetime (and that each quasar is a supermassive black hole).  If it is assumed that each of these bursts is due to the formation of a supermassive black hole via the collapse of a SMS, then Haehnelt's rate estimates can be used as estimates of the rate of SMS collapse. 
This rate is likely an overestimate of the SMS collapse rate because many SMBHs may have been formed via merger.   Haehnelt predicts that the integrated event rate through redshift $z$=$4.5$ ranges from $\sim 10^{-6}\, {\rm yr^{-1}}$ for $M$=$10^8M_{\odot}$ objects to $\sim 1\,{\rm yr^{-1}}$ for $M$=$10^6 M_{\odot}$ objects.  Thus, as in the case of Population III stars, a reasonable occurrence rate can be found for an observation (luminosity) distance of $50\,{\rm Gpc}$.

\subsection{GW Emission Mechanisms}
\label{section:sms-mech}

The GW emission mechanisms related to the collapse of SMSs are a subset of those discussed in the sections on AIC, SNe/collapsars, and Population III stellar collapse.  These include aspherical collapse, global rotational and fragmentation instabilities that may arise during the collapse/explosion and in the collapsed remnant (prior to black hole formation), and the ``ring--down'' of the remnant black hole.

\subsection{Numerical Predictions of GW Emission}
\label{section:sms-num}

The outcome of SMS collapse can only be determined with numerical, relativistic 3D hydrodynamics simulations.

Until recently, such simulations had only been published for nearly spherical collapse.  The spherical simulations of Shapiro and Teukolsky~\cite{shapiro-79} produced collapse evolutions that were nearly homologous.  In this case, the collapse time $\tau_{coll}$ is roughly the free--fall time at the horizon
\begin{equation}
\tau_{coll}=\left( \frac {R^3}{4\pi M}\right)^{1/2}=14\,{\rm s}[M/10^6 M_{\odot}]^{-1},
\end{equation}
The peak GW frequency $f_{GW}$=$\tau_{coll}^{-1}$ is then $10^{-2}\,{\rm Hz}$, if the mass of the star is $10^6 M_{\odot}$.  This is in the middle of LISA's frequency band of $10^{-4}$--$1\,{\rm Hz}$~\cite{thorne-95, folkner-98}.

The amplitude $h$ of this burst signal can be roughly estimated in terms of the star's quadrupole moment
\begin{eqnarray}
h &\leq& \epsilon \frac{2M^2}{Rd} \nonumber \\
  &\leq& \epsilon \, 1\times 10^{-18}\, [M/10^6 M_{\odot}][d/50\,{\rm Gpc}]^{-1}.
\label{equation:hsms}
\end{eqnarray}
Here $d$ is the distance to the star and $\epsilon \sim T/|W|$ is a measure of the star's deviation from spherical symmetry.  In this case, $\epsilon$ will be much less than one near the horizon, since the collapse is nearly spherical.

There are two possible aspherical collapse outcomes that can be discussed.  The first outcome is direct collapse to a SMBH.  In this case, $\epsilon$ will be on the order of one near the horizon.  Thus, according to equation~\ref{equation:hsms}, the peak amplitude of the GW burst signal will be $h_{pk}\sim 1 \times 10^{-18} [M/10^6 M_{\odot}][d/50\,{\rm Gpc}]^{-1}$.

\begin{figure}[h]
 \def\epsfsize#1#2{0.65#1}
 \centerline{\epsfbox{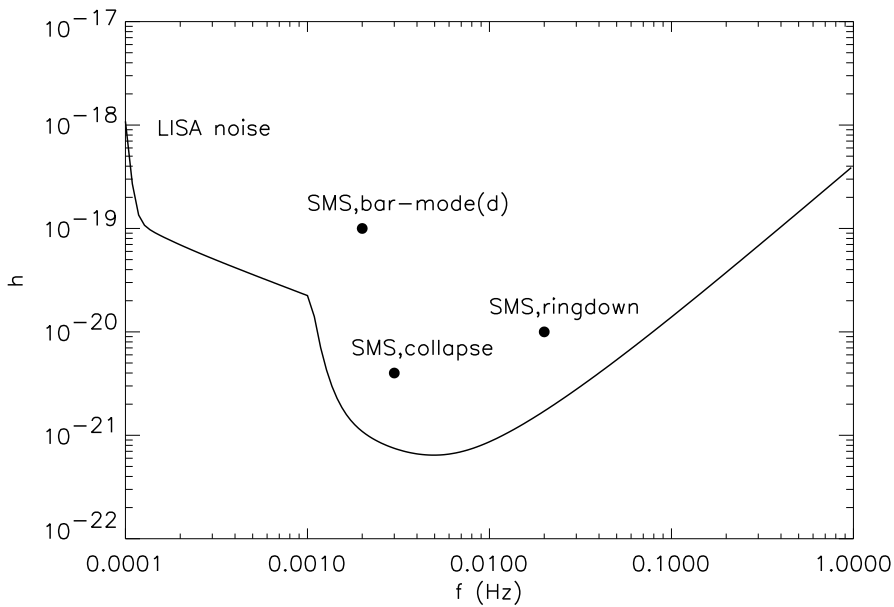}}
 \caption{\it  A comparison between the GW amplitude $h(f)$ for various sources and the LISA noise curve. See the text for details regarding the computations of $h$.  The SMS sources are assumed to be located at a luminosity distance of $50\,{\rm Gpc}$.  The bar--mode source is a dynamical bar--mode.}
 \label{figure:lisa}
\end{figure}

Alternatively, the star may encounter the dynamical bar mode instability prior to complete collapse.  Baumgarte and Shapiro~\cite{baumgarte-99} have estimated that a uniformly rotating SMS will reach $\beta \sim 0.27$ when $R/M$=$15$.  The frequency of the quasiperiodic gravitational radiation emitted by the bar can be estimated in terms of its rotation frequency to be
\begin{eqnarray}
f_{gw} &=& 2f_{bar}\sim 2 \left(\frac{GM}{R^3}\right)^{1/2} \nonumber \\
       &=& 2 \times 10^{-3} {\rm Hz}[M/10^6M_{\odot}]^{-1},
\label{equation:fbarsms}
\end{eqnarray}
when $R/M$=$15$.  The corresponding $h_{pk}$, again estimated in terms of the star's quadrupole moment, is
\begin{eqnarray}
h_{pk} &\leq& \frac{2M^2}{Rd} \nonumber \\
  &\leq& 1\times 10^{-19}\, [M/10^6 M_{\odot}][d/50\,{\rm Gpc}]^{-1}.
\label{equation:hbarsms}
\end{eqnarray}
The LISA sensitivity curve is shown in figure~\ref{figure:lisa} (see \cite{hughes-02b} for details on the computation of this curve; a mission time of $3\,{\rm yrs}$ has been assumed). The GW signal from this dynamical bar--mode could be detected with LISA. 

Shibata and Shapiro~\cite{shibata-02} have recently published a fully general relativistic, axisymmetric simulation of the collapse of a rapidly, rigidly rotating SMS.  They found that the collapse remained homologous during the early part of the evolution.  An apparent horizon does appear in their simulation, indicating the formation of a black hole.  Because of the symmetry condition used in their run, non--axisymmetric instabilities were unable to develop.

The collapse of a uniformly rotating SMS has been investigated with post-Newtonian hydrodynamics, in $3$+$1$ dimensions, by Saijo, Baumgarte, Shapiro, and Shibata~\cite{saijo-02}.  Their numerical scheme used a post-Newtonian approximation to the Einstein equations, but solved the fully relativistic hydrodynamics equations.  Their initial model was an $n$=$3$ polytrope.

\begin{figure}[h]
 \def\epsfsize#1#2{0.7#1}
 \centerline{\epsfbox{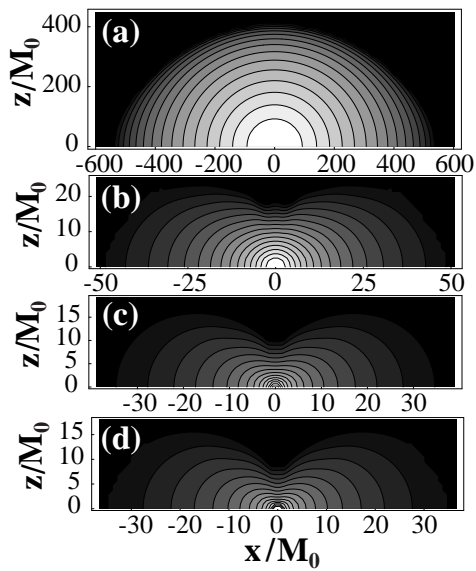}}
 \caption{\it Meridional plane density contours from the SMS collapse simulation of Saijo, Baumgarte, Shapiro, and Shibata~\cite{saijo-02}.  The contour lines denote densities $\rho$=$\rho_c \times d^{(1-i/16)}$, where $\rho_c$ is the central density.  The frames are plotted at ($t/t_{D}$, $\rho_c$, $d$)=(a)($5.0628\times 10^{-4}$, $8.254\times 10^{-9}$, $10^{-7}$), (b)($2.50259$, $1.225\times 10^{-4}$, $10^{-5}$), (c)($2.05360$, $8.328\times 10^{-3}$, $5.585\times10^{-7}$), (d)($2.50405$, $3.425\times 10^{-2}$, $1.357\times 10^{-7}$), respectively.  Here $t$, $t_{D}$, and $M_0$ are the time, dynamical time (=$\sqrt{R_e^{3}/M}$, where $R_e$ is the initial equatorial radius and $M$ is the mass), and rest mass.  (Figure 15 of~\cite{saijo-02}; used with permission.)}
 \label{figure:sms}
\end{figure}

The results of Saijo et al.\ indicate that the collapse of a uniformly rotating SMS is coherent (i.e., no fragmentation instability develops).  The collapse evolution of density contours from their model is shown in figure~\ref{figure:sms}. Although the work of Baumgarte and Shapiro~\cite{baumgarte-99} suggests that a bar instability should develop prior to BH formation, no bar development was observed by Saijo et al.  They use the quadrupole approximation to estimate a mean GW amplitude from the collapse itself: $h = 4 \times 10^{-21}$, for a $10^6 M_{\odot}$ star located at a distance of $50\,{\rm Gpc}$.  Their estimate for $f_{GW}$ at the time of BH formation is $3 \times 10^{-3}\,{\rm Hz}$.  This signal would be detectable with LISA (see figure~\ref{figure:lisa}).

Saijo et al.\ also consider the GW emission from the ringdown of the black hole remnant.
For the $l$=$m$=$2$ quasi-normal mode of a Kerr black hole with $a/M$=$0.9$, they estimate the characteristic frequency and amplitude of the emission to be $f_{GW}\sim 2\times 10^{-2}\,{\rm Hz}$ and $h \sim 1 \times 10^{-20}[(\triangle E_{GW}/M)/10^{-4}]^{1/2}$, for an $M=10^6 M_{\odot}$ source located at a luminosity distance of $50\,{\rm Gpc}$ (see~\cite{leaver-85, thorne-87, shibata-01} for details).  Here $\triangle E_{GW}/M$ is the radiated energy efficiency and may be $\lesssim 7\times 10^{-4}$~\cite{stark-85}.  This GW signal is within LISA's range of sensitivity (see figure~\ref{figure:lisa}).

\newpage


\section{Summary}
\label{section:summary}

It is hoped that as gravitational collapse simulations become more sophisticated, the historically widely varying estimates of the magnitude of GW emission from collapse may start to converge.  Steady progress in this field has been made in the last decade.  Some researchers have begun to use progenitor models produced with stellar evolution codes, which thus have more realistic angular momentum profiles, as starting points for collapse simulations~\cite{fryer-02}.  This reduces the need for collapse studies that include large surveys of the angular momentum parameter space.  Other progress made in the numerical study of collapse includes the use of realistic equations of state~\cite{fryer-02}, advanced neutrino transport and interaction schemes~\cite{janka-02c, lieben-02, rampp-02, thompson-02}, and the performance of 3D Newtonian~\cite{rampp-98, brown-01, fryer-02a} and improved 2D general relativistic simulations~\cite{shibata-00ptp}.

There is still much work to be done toward the goal of self--consistent, 3D general relativistic collapse simulations. Accurate progenitor modelling and collapse simulations must include the effects of magnetic fields, as they can significantly alter the amount of angular momentum and differential rotation present in collapsing stars.  Many of the more advanced studies, which include proper microphysics treatment and/or general relativistic effects, have been limited to axisymmetry.  Full 3D simulations are necessary to compute the characteristics of the GW emission from non--axisymmetric collapse phenomena.  Furthermore, simulations that follow both the collapse and the evolution of the collapsed remnant are necessary to consistently predict GW emission.  One benefit of long duration simulations is that they will facilitate the investigation of the effects of the envelope on any instabilities that develop in the collapsing core or remnant.  Of course, lengthy 3D simulations are computationally intensive.  This burden may be reduced by the use of advanced numerical techniques, including adaptive mesh refinement and parallel algorithms. 

The current numerical simulations of gravitational collapse indicate that interferometric observatories could detect GWs emitted by some collapse phenomena.  LIGO-I may be able to detect GWs from secular bar--mode instabilities in core--collapse SNe~\cite{lai-01} and magnetized tori surrounding black hole collapse remnants~\cite{putten-02}.  LIGO-II could observe GWs from dynamical bar--mode instabilities in AIC~\cite{liu-02} and core--collapse SNe~\cite{fryer-02}, and possibly from the fragmentation of very massive SNe cores that merge to form BHs~\cite{davies-02}.  LISA should be able to detect the collapse (and any bar--mode instabilities that develop during the collapse) of SMSs~\cite{baumgarte-99} and the ringdown of black hole remnants of collapsed Population III stars~\cite{fryer-02} and SMSs~\cite{saijo-02}.  These observations will provide unique information about gravitational collapse and its associated progenitors and remnants.

\clearpage
\newpage


\section{Acknowledgements}
\label{section:acknowledgements}

It is a pleasure to thank Paul Bradley, Adam Burrows, Harald Dimmelmeier, Chris Fryer, Alex Heger, Scott Hughes, Hui Li, Ewald M\"{u}ller, and Stuart Shapiro for helpful conversations and/or permission to reprint figures/movies from their published works.  I also gratefully acknowledge Ewald M\"{u}ller and a second referee for their beneficial reviews of this article.  This work was performed under the auspices of the US Department of Energy by the Los Alamos National Laboratory under contract W-7405-ENG-36.
\newpage

\bibliography{gc_gw}


\end{document}